%% file: rpv202_ep.tex
\documentclass[12pt]{article}
\usepackage{a4wide}
\usepackage[dvips]{graphics}

\newcommand{\roots}  {\ensuremath{\sqrt{s}}}
\newcommand{\tanb} {\ensuremath{\tan \beta }}

\newcommand{\gevc}{\mbox{GeV$/c$}}
\newcommand{\gevcc}{\mbox{GeV$/c^2$}}

\newcommand{\gev}{\mbox {GeV}}

\newcommand{\invpb}{\mbox{pb$^{-1}$}}

\newcommand{\q}{{\mathrm q}}

\newcommand{\slL}{\ensuremath{\mathrm{\tilde{l}_L}}}
\newcommand{\slR}{\ensuremath{\mathrm{\tilde{l}_R}}}
\newcommand{\seR}{\ensuremath{\mathrm{\tilde{e}_R}}}
\newcommand{\smuR}{\ensuremath{\mathrm{\tilde{\mu}_R}}}
\newcommand{\smuL}{\ensuremath{\mathrm{\tilde{\mu}_L}}}
\newcommand{\stauR}{\ensuremath{\mathrm{\tilde{\tau}_R}}}

\newcommand{\sbL}{\ensuremath{\mathrm{\tilde{b}_L}}}
\newcommand{\stL}{\ensuremath{\mathrm{\tilde{t}_L}}}
\newcommand{\sbR}{\ensuremath{\mathrm{\tilde{b}_R}}}
\newcommand{\suR}{\ensuremath{\mathrm{\tilde{u}_R}}}

\newcommand{\sdR}{\ensuremath{\mathrm{\tilde{d}_R}}}

\newcommand{\stone}{\ensuremath{\mathrm{\tilde{t}_1}}}
\newcommand{\sbone}{\ensuremath{\mathrm{\tilde{b}_1}}}

\newcommand{\emiss}{{\not \!\! E}}

\newcommand{\ALEPH}{{\tt ALEPH}}
\newcommand{\sq}{\ensuremath{\tilde{\mathrm q}}}
\newcommand{\sdq}{\ensuremath{\tilde{\mathrm d}}}
\newcommand{\suq}{\ensuremath{\tilde{\mathrm u}}}
\newcommand{\snu}{\ensuremath{\tilde{\nu}}}
\newcommand{\snue}{\ensuremath{\snu_\mathrm{e}}}
\newcommand{\snumu}{\ensuremath{\snu_{\mu}}}
\newcommand{\slep}{\ensuremath{\tilde{\ell}}}
\newcommand{\se}{\ensuremath{\tilde{\mathrm{e}}}}
\newcommand{\smu}{\ensuremath{\tilde{\mu}}}
\newcommand{\stau}{\ensuremath{\tilde{\tau}}}

\newcommand{\epem}{\ensuremath{\mathrm{e^+e^-}}}

\newcommand{\qq}{\ensuremath{\mathrm{q\bar{q}}}}

\newcommand{\PW}{\ensuremath{\mathrm{W}}}
\newcommand{\PZ}{\ensuremath{\mathrm{Z}}}
\newcommand{\Pe}{\ensuremath{\mathrm{e}}}

\newcommand{\nch}{\ensuremath{N_{\mathrm{ch}}}}
\newcommand{\mvis}{\ensuremath{M_{\mathrm{vis}}}}
\newcommand{\nlep}{\ensuremath{N_{\mathrm{lep}}}}
\newcommand{\elep}{\ensuremath{E_{\mathrm{lep}}}}
\newcommand{\enlep}{\ensuremath{E_{\mathrm{nonlep}}}}
\newcommand{\ehad}{\ensuremath{E_{\mathrm{had}}}}
\newcommand{\evis}{\ensuremath{E_{\mathrm{vis}}}}
\newcommand{\pmiss}{\ensuremath{p^\mathrm{miss}}}
\newcommand{\ptmiss}{\ensuremath{p^{\mathrm{miss}}_{\perp}}}
\newcommand{\pzmiss}{\ensuremath{p^{\mathrm{miss}}_\mathrm{z}}}
\newcommand{\phimix}{\ensuremath{\phi_\mathrm{mix}}}
\newcommand{\chiWW}{\ensuremath{\chi^2_\mathrm{WW}}}

\newcommand{\newc}{\newcommand}

\newc{\R}{$R$}
\newc{\charginom}{M_{\tilde \chi}^{+}}
\newc{\mue}{\mu_{\tilde{e}_{iL}}}
\newc{\mud}{\mu_{\tilde{d}_{jL}}}
\newc{\beq}{\begin{equation}}
\newc{\eeq}{\end{equation}}
\newc{\barr}{\begin{eqnarray}}
\newc{\earr}{\end{eqnarray}}
\newc{\ra}{\rightarrow}
\newc{\Da}{\Downarrow}
\newc{\lam}{\lambda}
\newc{\eps}{\epsilon}
\newc{\eq}[1]{(\ref{eq:#1})}
\newc{\eqs}[2]{(\ref{eq:#1},\ref{eq:#2})}
\newc{\etal}{{\it et al.}\ }
\newc{\Hbar}{{\bar H}}
\newc{\Ubar}{{\bar U}}
\newc{\Dbar}{{\bar D}}
\newc{\Ebar}{{\bar E}}
\newc{\eg}{{e.g.}\ }
\newc{\ie}{{i.e.}\ }
\newc{\nonum}{\nonumber}
\newc{\lab}[1]{\label{eq:#1}}
\newc{\lle}[3]{L_{#1}L_{#2}\Ebar_{#3}}
\newc{\lqd}[3]{L_{#1}Q_{#2}\Dbar_{#3}}
\newc{\udd}[3]{\Ubar_{#1}\Dbar_{#2}\Dbar_{#3}}
\newc{\slle}{LL\bar{E}}
\newc{\slqd}{LQ\bar{D}}
\newc{\sudd}{\bar{U}\bar{D}\bar{D}}
\newc{\dpr}[2]{({#1}\cdot{#2})}
\newc{\rpv}{{\not R_p}}

\newc{\rpvm}{{\not \! R_p}}
\newc{\rp}{$R_p$}
\newc{\gsim}{ \,  \scriptstyle{\stackrel{>}{\sim}}\displaystyle \, }
\newc{\lsim}{ \,  \scriptstyle{\stackrel{<}{\sim}}\displaystyle \, }

\begin{document}
\begin{titlepage} 

\begin{center}
\rm EUROPEAN ORGANIZATION FOR NUCLEAR RESEARCH (CERN)
\end{center}

\begin{flushright}
CERN-EP/2000-132\\
25 October, 2000\\
\end{flushright}

\vspace{2cm}

\begin{center}
\LARGE{Search for R-Parity Violating Decays of Supersymmetric Particles
in ${\mathrm e}^+{\mathrm e}^-$ Collisions at Centre-of-Mass Energies
from 189~GeV to 202~GeV

\vspace{1cm}}

{\large The \ALEPH\ Collaboration\footnote{See next page for a list of
the authors}}
\end{center}

\thispagestyle{empty}


\vspace{2cm}
\begin{abstract}
\vspace{.5cm} Searches for the production of supersymmetric
particles under the assumption that R-parity is violated via a single
dominant $LL{\bar E}$, $LQ{\bar D}$ or ${\bar U} {\bar D} {\bar D}$
coupling were performed. These use the data collected by the \ALEPH\
detector at LEP at centre-of-mass energies from $188.6$ to
$201.6~\gev$.  The numbers of candidate events observed in the data are
consistent with Standard Model expectations. Upper limits on the
production cross sections and lower limits on the masses of charginos,
sleptons, squarks and sneutrinos are derived.
\end{abstract}

\vfill

\centerline{\large \em (Submitted to European Physical Journal C) }

\end{titlepage}

\newpage
\include{authb}

\section{Introduction}

Minimal supersymmetric extensions of the Standard Model (MSSM)~\cite{MSSM} 
usually make the assumption that R-parity,
$R_p=-1^{3B+L+2S}$, is conserved~\cite{fayet}, 
where $B$ denotes the baryon number, $L$ the lepton number and $S$ the spin 
of a field. The conservation of R-parity is not required theoretically and 
models in which R-parity is violated can be constructed 
which are compatible with existing experimental constraints. 

\begin{figure}
\begin{center}
\resizebox{0.75\textwidth}{!}{
  \resizebox{0.25\textwidth}{!}{
        \includegraphics{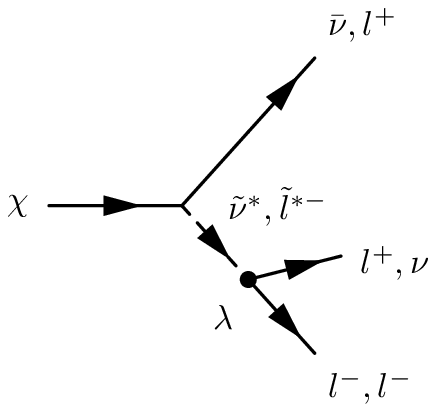}}
  \resizebox{0.25\textwidth}{!}{
        \includegraphics{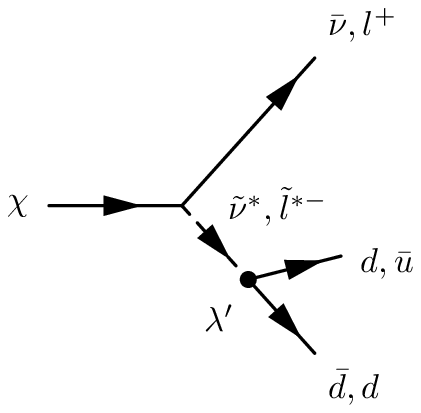}}
  \resizebox{0.25\textwidth}{!}{
        \includegraphics{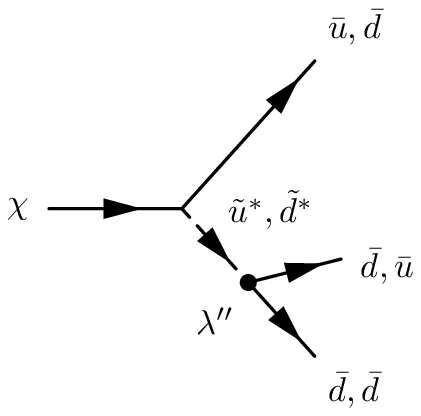}}
}
\resizebox{0.25\textwidth}{!}{
        \includegraphics{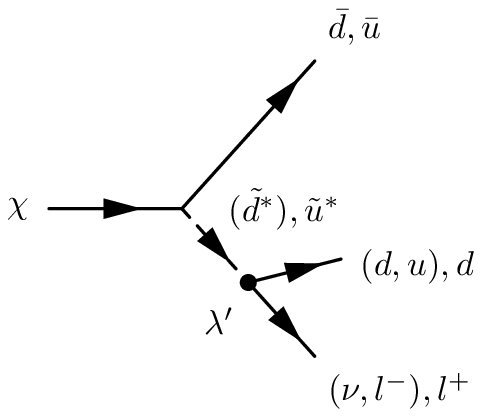}}
\resizebox{0.8\textwidth}{!}{
  \resizebox{0.25\textwidth}{!}{
        \includegraphics{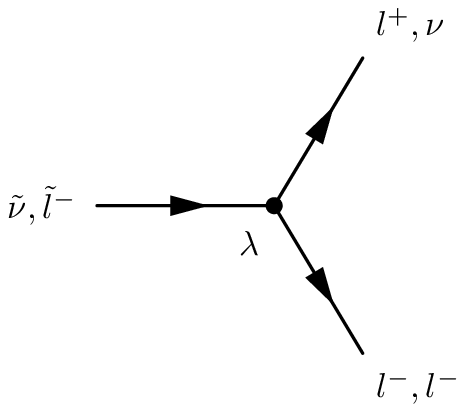}}
  \resizebox{0.25\textwidth}{!}{
        \includegraphics{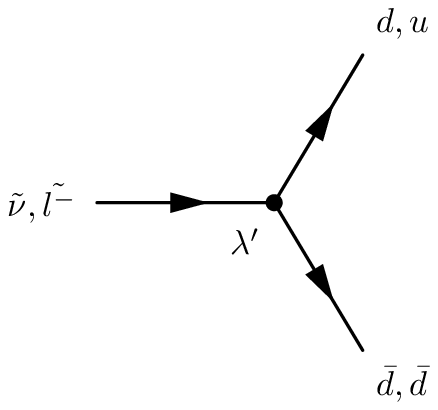}}
  \resizebox{0.25\textwidth}{!}{
        \includegraphics{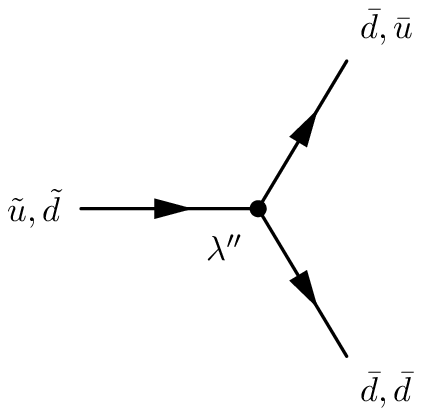}}
}
\resizebox{0.25\textwidth}{!}{
  \includegraphics{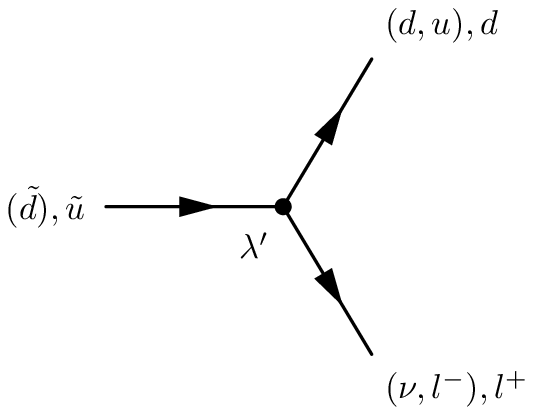}}
\end{center}
\caption[.]{ \small {\em Direct} R-parity violating decays of
supersymmetric particles via the
$\lambda$, $\lambda'$ and $\lambda''$ couplings. The points mark the
R-parity violating vertex in the decay.}
\label{dec.examples}
\end{figure}

The R-parity violating terms of the superpotential considered here
are~\cite{rpsuper}
\begin{equation}
W_{\rpv} =  \lam_{ijk}\lle{i}{j}{k}+ \lam'_{ijk}\lqd{i}{j}{k}+
            \lam''_{ijk}\udd{i}{j}{k},
\label{eqrpv}
\end{equation}
where $\Dbar,\Ubar$ ($\Ebar$) are the down-like and up-like quark
(lepton) singlet superfields, and $Q$ ($L$) is the quark (lepton)
doublet superfield respectively; $\lambda$, $\lambda'$ and $\lambda''$
are Yukawa couplings and $i,j,k=1,2,3$ are generation indices. The
presence of such R-parity violating terms imply that the lightest
supersymmetric particle (LSP) is no longer stable and that sparticles
can be produced singly. The sparticle decays which proceed directly
to standard model particles are called {\it direct} decays. Decays in
which the sparticle first decays, conserving R-parity, to the lightest
neutralino are referred to as {\it indirect} decays. Both cases are
illustrated in Fig.~\ref{dec.examples}. Other cascade decays are possible
but not considered in the following.

In this paper a new search for the resonant production of single 
sneutrinos decaying indirectly is presented. In addition, previously
reported searches for both direct and indirect decays of pair produced
sparticles at 183 GeV~\cite{183paper} are extended and applied to new
data at higher energies.  In particular, new selections for indirect
decays of sleptons and squarks via the $\slqd$ operator are developed.
Table \ref{poss_decays} summarises the possible decays and indicates
those addressed in this paper.  Other collaborations at LEP have
published similar searches at lower energies~\cite{DELPHI:LLE183,
OPAL:UDD_lep2, OPAL:Sfermion, L3:RPVsnu, L3:RPV183}.

The following assumptions are made throughout:
\begin{itemize}

\item{All three terms in Equation~(\ref{eqrpv}) are addressed, however
only one term for a specific set of indices ($i$, $j$ and $k$) is
considered non zero. Unless otherwise stated the derived limits
correspond to the choice of indices for the coupling giving the worst
limit.}

\item{The lifetime of the sparticles can be neglected, \ie the mean
flight path is less than $1$~cm.}

\item{Results are interpreted within the framework of the
MSSM. Gaugino mass unification at the electroweak scale is assumed,
giving the condition $M_1={5\over3}M_2\tan^2\theta_W$.}

\item{For the case of the charginos and neutralinos, only large values
 of the universal scalar mass $m_0$ are considered; this implies the 
$direct$ decays of the lightest chargino and the next-to-lightest 
neutralino are suppressed. It also implies three-body decay kinematics 
for the lightest neutralino.  }

\end{itemize}

The search results reported here use data collected by the \ALEPH\
detector in 1998 and 1999 from  ${\mathrm e}^+{\mathrm e}^-$ 
collisions at centre-of-mass energies
between $188.6~\gev$ and $201.6~\gev$.  The total data sample used
corresponds to an integrated recorded luminosity of
$173.6~\invpb$ at $188.6~\gev$, 
$29.0~\invpb$ at $191.6~\gev$, 
$80.1~\invpb$ at $195.5~\gev$,
$85.9~\invpb$ at $199.5~\gev$ and
$41.9~\invpb$ at $201.6~\gev$.

This paper is organised as follows: 
after a brief description of the ALEPH detector in 
Section~\ref{aleph.detector},
the Monte Carlo samples used
for signal and background generation are detailed in Section
\ref{samples}. Sections \ref{lle}, \ref{lqd} and \ref{udd} give the
results and interpretations for each of the R-parity violating
couplings, and finally Section \ref{conclusions} gives a summary of the
results.

\begin{table}
\caption[.]{\small For each sparticle the table lists 
whether the decay mode is searched for ($\bullet$), possible but not
considered ($\diamond$), or not possible ($\times$).  Those processes
marked with $\dagger$ were not considered in the 183 GeV results
\cite{183paper}.}
\label{poss_decays}
\begin{center}
\renewcommand{\arraystretch}{1.1}
\begin{tabular}{|c|c|c|c|c|c|c|}
\hline
    & \multicolumn{2}{c|}{$\slle$} & \multicolumn{2}{c|}{$\slqd$} &
      \multicolumn{2}{c|}{$\sudd$} \\
    \cline{2-7}
    & Direct & Indirect & Direct & Indirect & Direct & Indirect \\
\hline
$\chi^\pm$ & 
$\diamond$ & $\bullet$ & $\diamond$ & $\bullet$ & 
$\diamond$ & $\bullet$ \\

$\chi'$ & 
$\diamond$ & $\bullet$ & $\diamond$ & $\bullet$ & $\diamond$ & $\bullet$ \\

$\se,\smu,\stau$ & 
$\bullet$ & $\bullet$ & $\bullet$($\slL$) & $\bullet^\dagger$ & $\times$ &
$\bullet$ \\

$\snue,\snumu,\snu_\tau$ & 
$\bullet$ & $\bullet$ & $\bullet$ & $\bullet^\dagger$ & $\times$ & $\bullet$ \\

$\suq$ &
$\times$ & $\bullet$ & $\bullet$ & $\bullet^\dagger$ & $\bullet(\suR)$ & $\bullet$ \\

$\sdq$ &
$\times$ & $\bullet$ & $\bullet$ & $\bullet^\dagger$ & $\bullet(\sdR)$ & $\bullet$ \\
\hline
\end{tabular}
\renewcommand{\arraystretch}{1.}
\end{center}
\end{table}

\section{\label{aleph.detector}The ALEPH Detector}

The \ALEPH\ detector is described in detail in
Ref.~\cite{bib:detectorpaper}. An account of the performance of the
detector and a description of the standard analysis algorithms can be
found in Ref.~\cite{bib:performancepaper}. Here, only a brief
description of the detector components and the algorithms relevant for
this analysis is given.

The trajectories of charged particles are measured with a silicon
vertex detector, a cylindrical drift chamber, and a large time
projection chamber (TPC). The central detectors are immersed in a
1.5~T axial magnetic field provided by a superconducting solenoidal
coil.  The electromagnetic calorimeter (ECAL), placed between the TPC
and the coil, is a highly segmented sampling calorimeter which is used
to identify electrons and photons and to measure their energies.  The
luminosity monitors extend the calorimetric coverage down to 34~mrad
from the beam axis.  The hadron calorimeter (HCAL) consists of the
iron return yoke of the magnet instrumented with streamer tubes. It
provides a measurement of hadronic energy and, together with the
external muon chambers, muon identification.  The calorimetric and
tracking information are combined in an energy flow algorithm which
gives a measure of the total energy, and therefore the missing energy,
with an uncertainty of $(0.6\sqrt{E}+0.6)$~GeV.

Electron identification is primarily based upon the matching between
the measured momentum of the charged track and the energy deposited in
the ECAL. Additional information from the shower profile in the ECAL
and the measured rate of specific ionisation energy loss in the TPC
are also used.  Muons are separated from hadrons by their
characteristic pattern in HCAL and the presence of associated hits in
the muon chambers.

\section{\label{samples}Monte Carlo Samples and Efficiencies}

The signal topologies were simulated using the {\tt SUSYGEN} Monte
Carlo program~\cite{susygen} modified as described in 
Ref.~\cite{183paper}. The events were subsequently passed through either
a full simulation or a faster simplified simulation of the ALEPH
detector. Where the fast simulation was used a subselection of these
were also passed through the full simulation to verify the accuracy of
the fast simulation.

Samples of all major backgrounds were generated and passed through the
full simulation, corresponding to at least 10 times the collected
luminosity in the data. The {\tt PYTHIA} generator~\cite{pythia} was
used to produce $\qq$ events and four-fermion final states from
$\PW\Pe\nu$, $\PZ\PZ$ and $\PZ\Pe\Pe$, with a vector-boson invariant
mass cut of $0.2~\gevcc$ for $\PZ\PZ$ and $\PW\Pe\nu$, and $2~\gevcc$
for $\PZ\Pe\Pe$. Pairs of W bosons were generated with {\tt
KORALW}~\cite{koralw}. The {\tt KORALW} cross sections were adjusted
to agree with the most recent theoretical
calculations~\cite{wwcs_correction}. Pair production of leptons was
simulated with {\tt UNIBAB}~\cite{unibab} (electrons) and {\tt
KORALZ}~\cite{koralz} (muons and taus). The $\gamma\gamma \rightarrow
\mathrm{f\bar{f}}$ processes were generated with {\tt
PHOT02}~\cite{phot02}.

The selections were optimised to give the minimum expected $95\%$
C.L. excluded cross section in the absence of a signal for masses
close to the high end of the expected sensitivity. Selection
efficiencies were determined as a function of the SUSY particle masses
and the generation structure of the R-parity violating couplings
$\lambda_{ijk}$, $\lambda'_{ijk}$ and $\lambda^{''}_{ijk}$.

The cross section limits were evaluated at the highest centre-of-mass
energy. Where data taken at a range of centre-of-mass energies
contributed to the exclusions the data were weighted with the expected
evolution of the cross-section with $\roots$.

The systematic uncertainties on the selection efficiencies are of
order of 4--5$\%$ and are dominated by the statistical uncertainty of
the Monte Carlo signal samples, with small additional contributions
from lepton identification and energy flow reconstruction. They were
taken into account by reducing the selection efficiencies by one
standard deviation of the statistical error.  

When setting the limits, background subtraction was performed for two-
and four-fermion final states according to the prescription given in
Ref. \cite{PDG}. To take into account the uncertainties on the
background estimates, the amount of background subtracted is
reduced. For two-fermion processes it was reduced by its statistical
error.  The contribution from WW and ZZ processes were reduced by the
statistical error added in quadrature with $1\%$ of its estimate.  The
components from $\PW\Pe\nu$ and $\PZ\Pe\Pe$ processes were reduced by
$20\%$ of their estimate.  No background is subtracted for the
$\gamma\gamma \rightarrow \mathrm{f\bar{f}}$ processes.

\section {\label{lle} Decays via a dominant \boldmath $ \slle$ coupling}

Under the assumption of a dominant $\slle$ coupling, the decay
topologies can consist of as little as two acoplanar leptons in the
simplest case (direct slepton decay or single resonant sneutrino
production), or they may consist of as many as six leptons plus four
neutrinos in the most complicated case (indirect chargino decay).  In
addition to the purely leptonic topologies, the MSSM cascade decays of
charginos into lighter neutralinos may produce multi-jet and
multi-lepton final states. No direct decays are possible for the
squarks.

The absolute lower limit on the mass of the lightest neutralino of
$23~\gevcc$ obtained in Ref.~\cite{LLEpaper}, which is valid for any
choice of $\mu$, $M_2$, $m_0$ and generational indices ($i$, $j$ and
$k$), is used to restrict the range of neutralino mass considered for
the indirect decays.

The various selections addressing the above topologies, the expected
backgrounds, and the numbers of candidates selected in the data at
$\roots=188.6$--$201.6~\gev$ are summarised in Table~\ref{topslle}.
Details of the ``6 Leptons + $\emiss$'', the ``4 Leptons + $\emiss$''
and the ``4 Leptons'' analyses are given in Ref.~\cite{LLEpaper}.  The
``Acoplanar Leptons'' and ``Leptons and Hadrons'' selections are
described in Ref.~\cite{183paper}, although the ``Leptons and
Hadrons'' has been updated for the increased centre-of-mass energy as
described in section~\ref{challe}. Wherever the ``Acoplanar Leptons''
selection is used the expected combination of final state flavours (e,
$\mu$ and $\tau$) for each process is used to set the exclusion.

\begin{table}
\caption[.]{ \small   
The observed numbers of events in the data and the corresponding Standard Model
background expectations for the $\slle$ selections.}
\label{topslle}
\setlength{\tabcolsep}{0.12cm}
\begin{center}
\begin{tabular}{|c|cc|cc|cc|cc|cc||cc|}
\hline
\multicolumn{1}{|r|}{$\roots$ (GeV)} &
\multicolumn{2}{c|}{188.6} & \multicolumn{2}{c|}{191.6} &
\multicolumn{2}{c|}{195.5} & \multicolumn{2}{c|}{199.5} &
\multicolumn{2}{c||}{201.6} &\multicolumn{2}{c|}{All} \\
Selection & 
Data & SM & Data & SM &  Data & SM & Data & SM & Data & SM & Data & SM \\
\hline
\hline

Leptons and& 
10 & 7.8 & 1  &  1.4 &    4  &  4.0&     5 &   4.5 &    0  &  2.1 &
20 & 20\\
Hadrons &&&&&&&&&&& &\\
\hline

6 Leptons &
2 &  1.0 &  1 &  0.2 &  0 &  0.4 &  0 &  0.5 &  0 &  0.2 &   3 &  2.2 \\
+ $\emiss$ &&&&&&&&&&&& \\
\hline

4 Leptons &
4 &  4.6 &  1 &  0.8 &  1 &  2.0 &  4 &  2.8 &  1 &  1.7 &  11 &
12 \\
+ $\emiss$ &&&&&&&&&&&& \\
\hline

 $llll$ &
3   & 5.1   & 1  &  0.7 &    3  &  1.7&     8 &   2.3 &    0  &  1.3 &
15 & 11\\
&&&&&&&&&&&& \\
\hline

$ll\tau\tau$&
2   & 2.0   &  0  &  0.3 &    2  &  0.9&     0 &   0.9 &    0  &  0.4 &
4 & 4.6\\
&&&&&&&&&&&& \\
\hline

$\tau\tau\tau\tau$ &
5 &  4.2 &  1 &  0.7 &  1 &  1.9 &  5 &  2.1 &  0 &  1.1 &  12 &
10 \\
&&&&&&&&&&&& \\
\hline

Acoplanar&
192 & 211 & 22 &  34 & 93 &  87 &100 &  94 & 41 &  45 & 448
& 471 \\ 
 Leptons  &&&&&&&&&&&& \\
\hline
\end{tabular}
\end{center}
\end{table}

\subsection{Single resonant sneutrino production}

\begin{figure}
\begin{center}
\resizebox{0.22\textwidth}{!}{
  \includegraphics{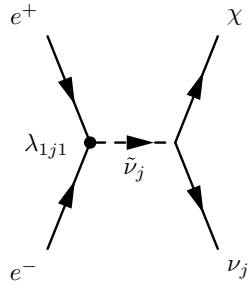}
}
\caption[.]{\small Single sneutrino production via the $\lambda_{1j1}$
coupling and subsequent {\em indirect} decay.}
\label{fig:single_snu}
\end{center}
\end{figure}

 Single resonant production of sneutrinos~\cite{resonance} can occur
for the specific couplings $\lambda_{121}$ and $\lambda_{131}$. The
sneutrino may decay indirectly through the diagram shown in
Fig.~\ref{fig:single_snu} or directly to $\epem$.  Since the
production cross-section is a function of $|\lam_{1j1}|^2$, limits can
be set on the magnitude of $\lam_{1j1}$ as a function of the sneutrino
mass. The best sensitivity is obtained for the case where the
sneutrino is produced exactly on shell, $\roots=M_{\snu}$. For
centre-of-mass energies above $M_{\snu}$ initial state radiation from
the $\epem$ system allows a radiative return to the sneutrino
resonance. There is also some sensitivity for $M_{\snu}>\roots$ via
the production of a virtual sneutrino.

\begin{figure}
\begin{center}
\resizebox{\textwidth}{!}{
  \includegraphics{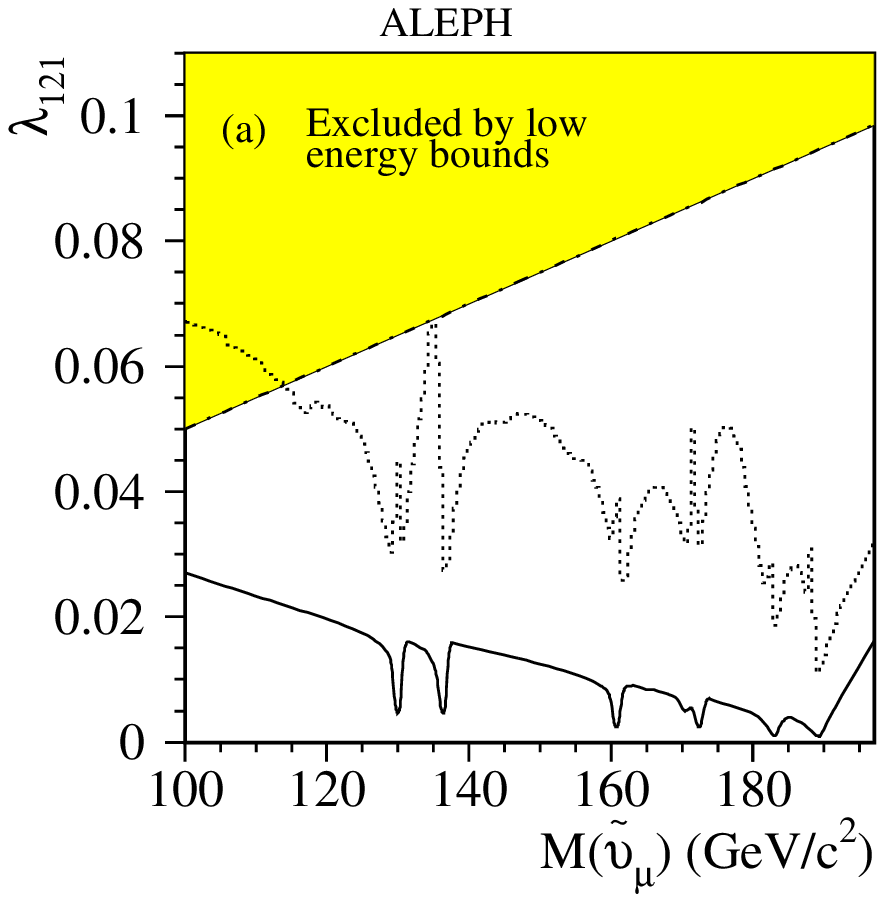}
  \includegraphics{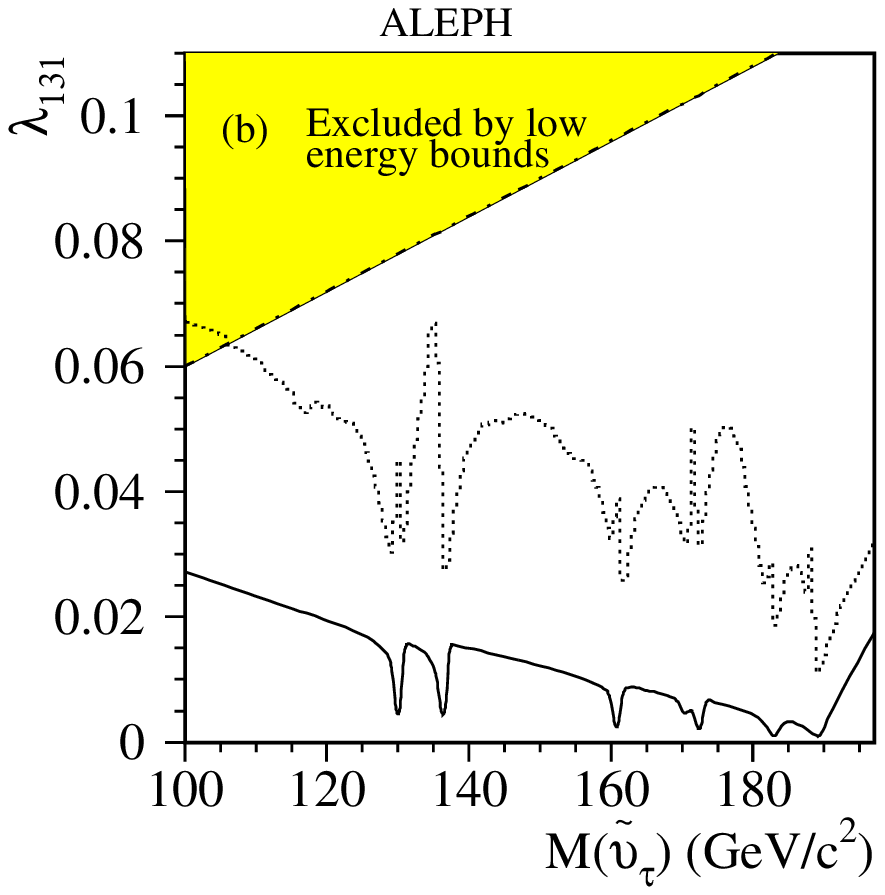}
}
\caption[.]{ \label{fig:single_snu_excl}  \small 
Plots (a) and (b) show the $95\%$ C.L. upper limits on the value of
the R-parity violating couplings, $\lambda_{121}$ and $\lambda_{131}$,
as a function of sneutrino mass for single sneutrino production and
indirect decays (solid curve); the limits are shown for the neutralino
mass giving the worst limit. For comparison the limit, assuming
$100\%$ branching ratio, for the direct decays to $\epem$ is also
shown (dotted histogram). Assuming that $M(\tilde{\nu}_{j})=
M(\tilde{e}_{R})$, the shaded region is excluded by (a) charged
current universality and (b) $R_{\tau}$. The exclusions are evaluated
at $\mu=-200~\gevcc$ and $\tanb=2$.  }
\end{center}
\end{figure}

Since the final state consists of two leptons and two neutrinos the
``Acoplanar Leptons'' selection is used to select these events.
Figure~\ref{fig:single_snu_excl} shows the excluded values of
$\lambda_{121}$ and $\lambda_{131}$ as a function of the mass of the
sneutrinos; all data taken in the range $\roots=130$ to $189$ GeV
are used. The data and background numbers for the lower energies are
given in Ref.~\cite{183paper,LLEpaper}.  Also shown are the results for the
direct decays from electroweak fits~\cite{tomalin} and the exclusion
from low energy measurements~\cite{low_energy_snu}.

\subsection {Charginos and neutralinos decaying via \boldmath $\slle$}
\label{challe}

Depending on the masses of the gauginos and on the lepton flavour
composition in the decay, the indirect decays of charginos to
neutralinos and of heavier neutralinos to lighter neutralinos
populate different regions in track multiplicity, visible mass and
leptonic energy. For this reason three different subselections were
developed~\cite{183paper}, covering topologies with large leptonic
energies and at least two jets (Subselection~I), topologies with small
multiplicities and large leptonic energy fractions (Subselection~II),
and topologies with a moderate leptonic energy fraction
(Subselection~III). The combination of the three sub-selections is
defined as the ``Leptons and Hadrons'' selection.  The complete set of
cuts, updated for $\roots>184~\gev$ is shown in Table~\ref{hlcuts}.

\begin{figure}
\begin{center}
\resizebox{\textwidth}{!}{
  \includegraphics{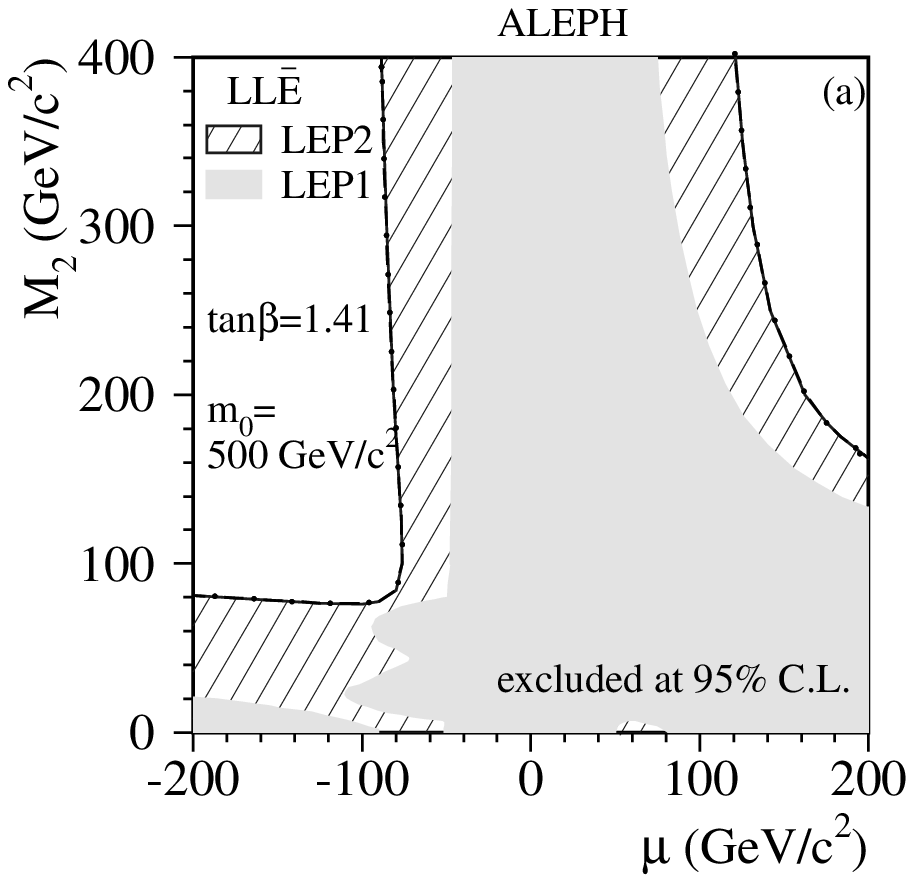}
  \includegraphics{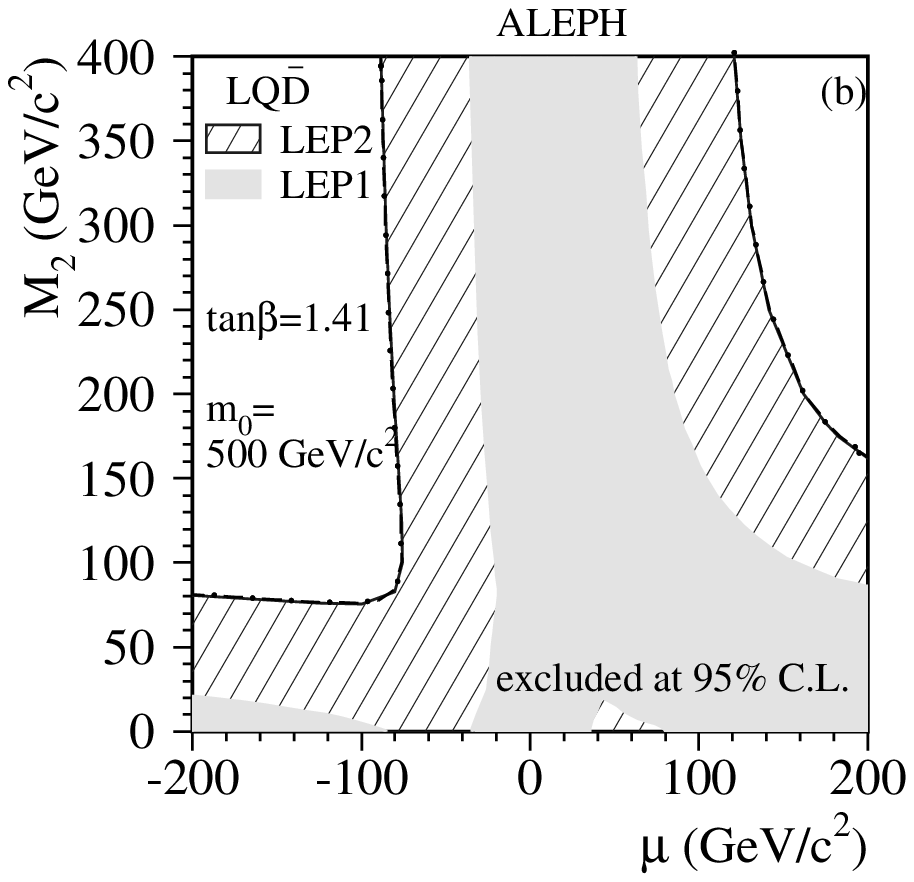}
}
\resizebox{0.5\textwidth}{!}{
  \includegraphics{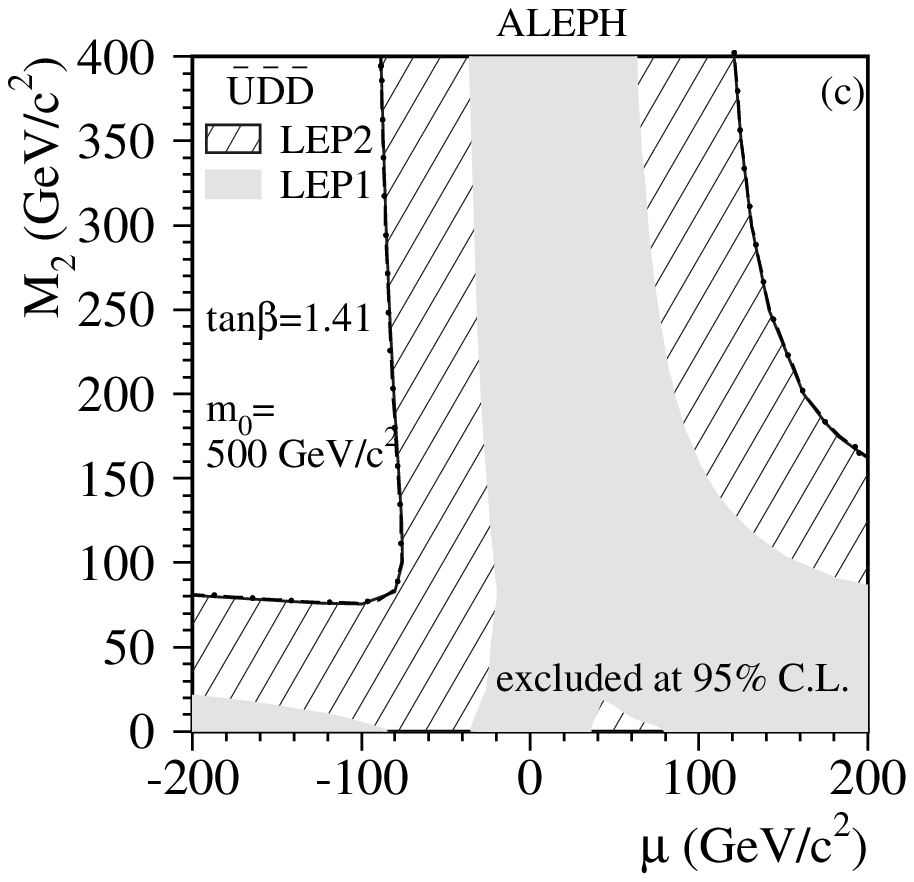}
}
\caption[.]{  \label{mum2lims}{\small Regions in the $(\mu,M_2)$
plane excluded at $95\%$ C.L. at $\tanb=1.41$ and $m_0=500~\gevcc$ for
the three operators.}}
\end{center}
\end{figure}

Interpreting the results, shown in Table~\ref{topslle}, in the
framework of the MSSM, 95$\%$ C.L. exclusion limits are derived in the
($\mu$,$M_2$) plane and shown in Fig.~\ref{mum2lims}(a) for large
scalar masses $m_0=500~\gevcc$. The corresponding lower limit on the
mass of the lightest chargino is essentially at the kinematic limit
for pair production. 

The searches for the lightest and second lightest
neutralino do not extend the excluded region in the
($\mu$,$M_2$) plane beyond that achieved with the chargino 
search alone.

\begin{table}[t]
\caption{\small \label{hlcuts}The list of cuts for the
``Leptons and Hadrons'' selection, which is used for charginos and
squarks decaying indirectly via the $\slle$ operator. The event variables are
defined in Ref.~\cite{183paper}.}
\begin{center}
\begin{tabular}{|c|c|c|} \hline
 Subselection~I & Subselection~II & Subselection~III\\
\hline \hline
$\nch{} \geq 5$ & $5 \leq \nch{} \leq 15$ & $\nch{} \geq 11$\\
$\mvis{} > 25~\gevcc$ & $20~\gevcc < \mvis{} < 75\%\sqrt{s}$ &
  $55\%\sqrt{s} < \mvis{} < 80\%\sqrt{s}$\\
\hline
$\ptmiss>3.5\%\sqrt{s}$ & $\ptmiss>2.5\%\sqrt{s}$ &
  $\ptmiss>5\%\sqrt{s}$ \\
$|\pzmiss| < 20~\gevc{}$ &  &
   $\nch^{\mathrm{jet}} \geq 1$\\
\hline
 & $y_3 > 0.009$ & $y_3>0.025$\\
 & $y_4 > 0.0026$ & $y_4>0.012$\\
$y_5 > 0.006$ &  & $y_5 > 0.004$ \\
 & & $T < 0.85$\\
\hline
$\nlep \geq 1$ & $\nlep \geq 1$ & $\nlep \geq 1$ \\
$\enlep < 50\%\sqrt{s}$ & $\enlep < 50\%\sqrt{s}$ & \\
$\ehad < 28\%\evis$ & $\ehad < 22\%\elep$ & $\elep > 20\% \ehad$\\
\hline
\multicolumn{3}{|c|}
{$ \chiWW > 3.8 $} \\
\hline
\end{tabular}
\end{center}
\end{table}

\subsection{Squarks decaying via \boldmath $LL{\bar E}$ }
\begin{figure}
\begin{center}
\resizebox{\textwidth}{!}{
  \includegraphics{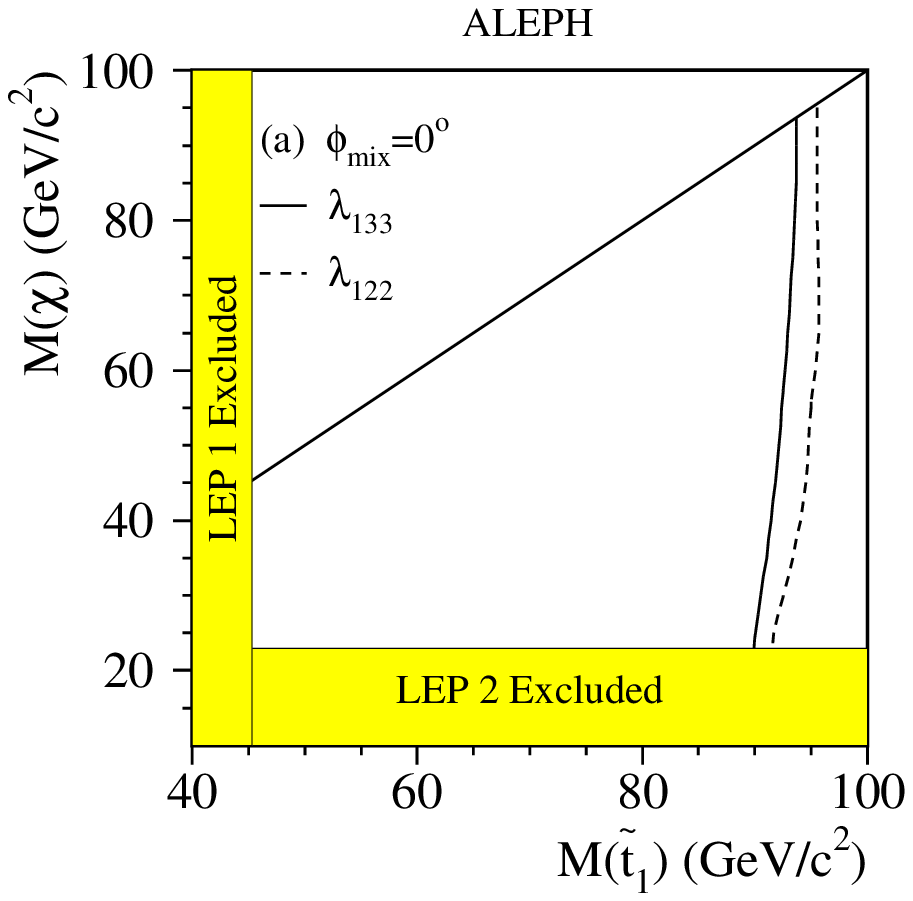}
  \includegraphics{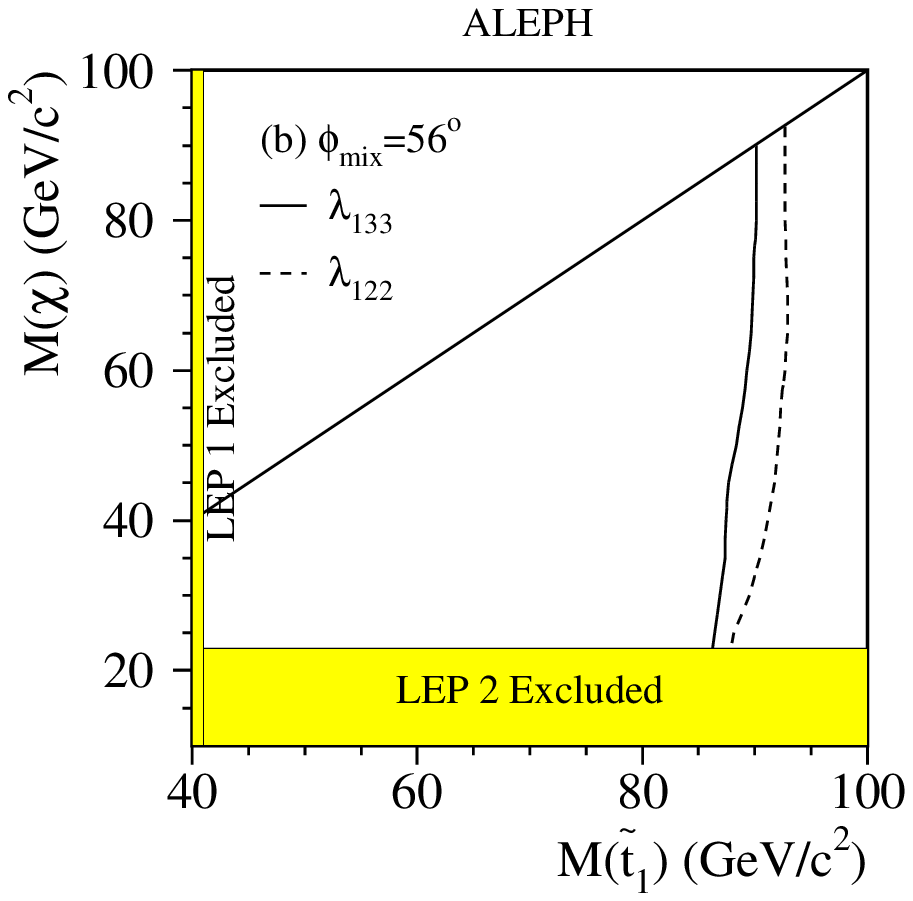}
}
\resizebox{\textwidth}{!}{
  \includegraphics{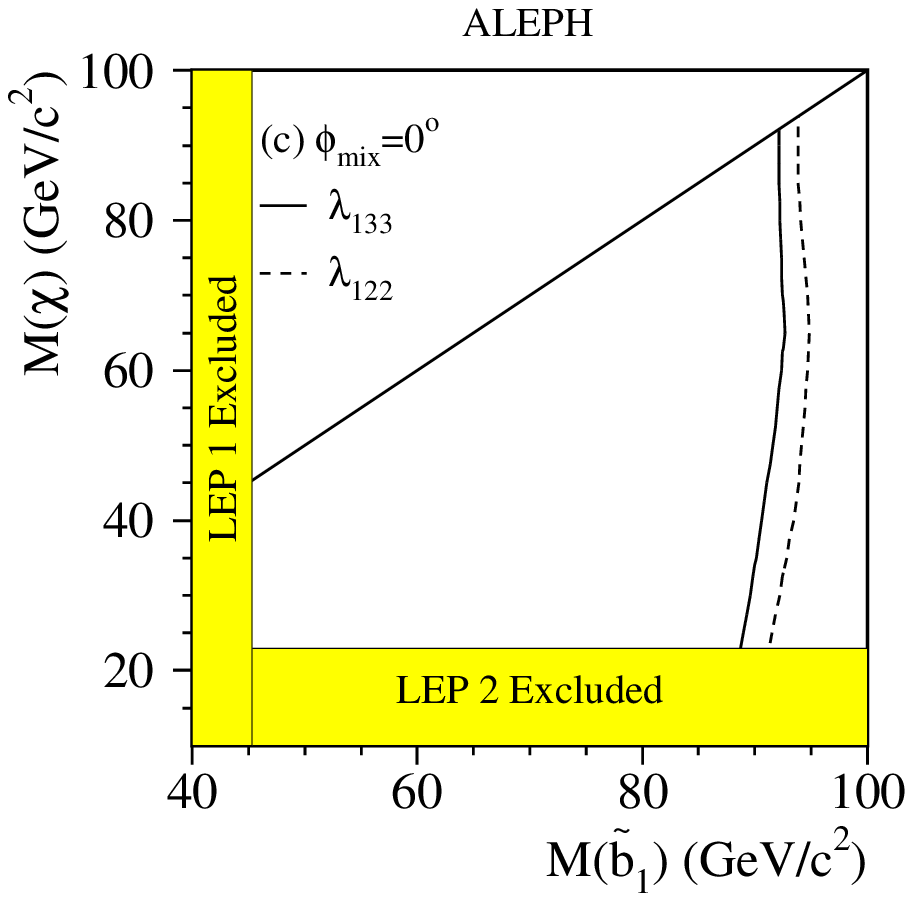}
  \includegraphics{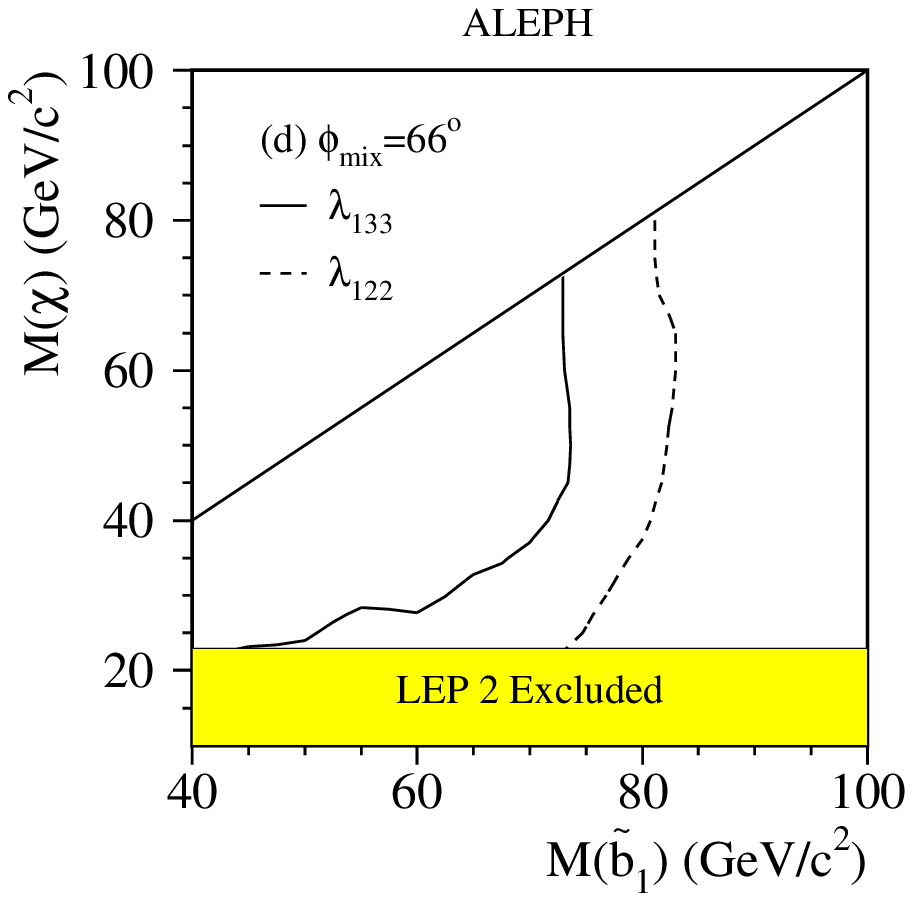}
}
\caption[.]{  \label{lle_stops}{\small The $95\%$ C.L. limits in the
($M_{\chi}$, $M_{\stone}$) and ($M_{\chi}$, $M_{\sbone}$) planes for
indirect decays via the $\slle$ couplings 
$\lambda_{122}$ and $\lambda_{133}$ are shown
for no mixing ($\phimix=0^\circ$) and for $\phimix=56^\circ,66^\circ$
for stops and sbottoms, respectively. The LEP 2 exclusion corresponds 
to the absolute limit on $M_{\chi}$.}}
\end{center}
\end{figure}

  Although squarks cannot decay directly with an $\slle$ coupling,
they may decay indirectly to the lightest neutralino. This topology is
searched for by means of the ``Leptons and Hadrons'' selection.  The
$95\%$ C.L. squark mass limits are presented as functions of
$M_{\chi}$ in Fig.~\ref{lle_stops} for the case of $\stone$ and
$\sbone$ squarks. The following limits upon the right-handed squarks
can be derived: $M_{\suR}>90~\gevcc$ and $M_{\sdR}>89~\gevcc$ for any
$\lam_{ijk}$.

\subsection{Sleptons decaying via \boldmath $LL{\bar E}$ }

\begin{figure}[t]
\begin{center}
\resizebox{\textwidth}{!}{
  \includegraphics{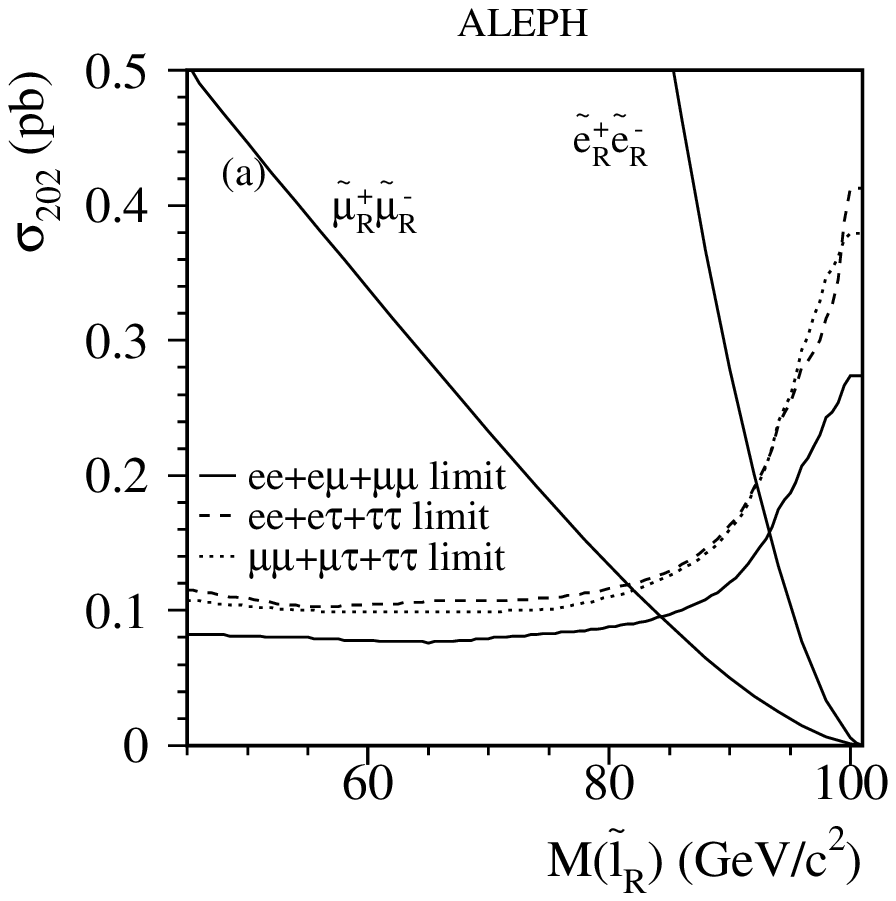}
  \includegraphics{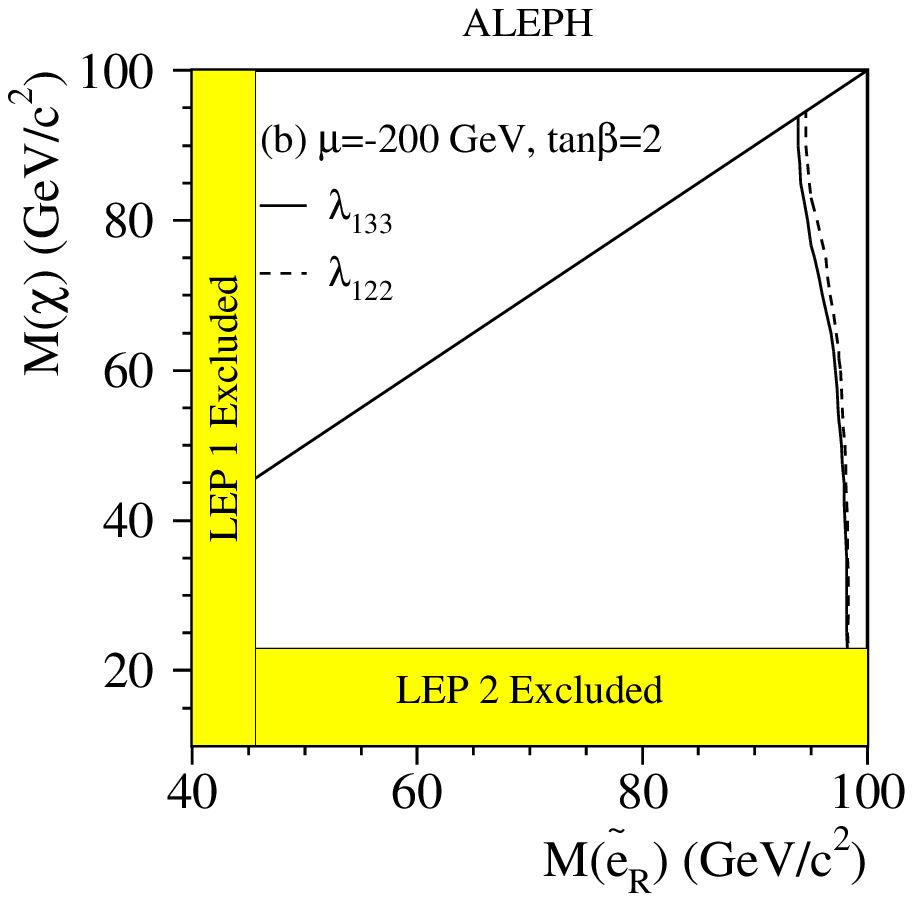}
}
\resizebox{\textwidth}{!}{
  \includegraphics{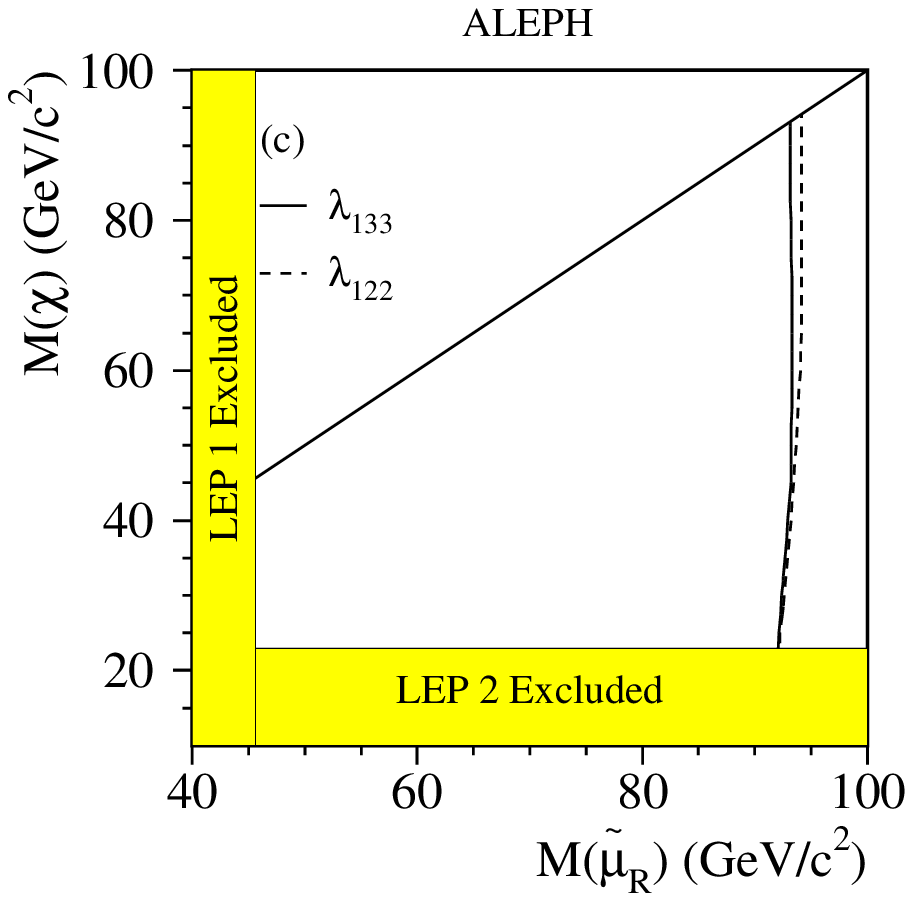}
  \includegraphics{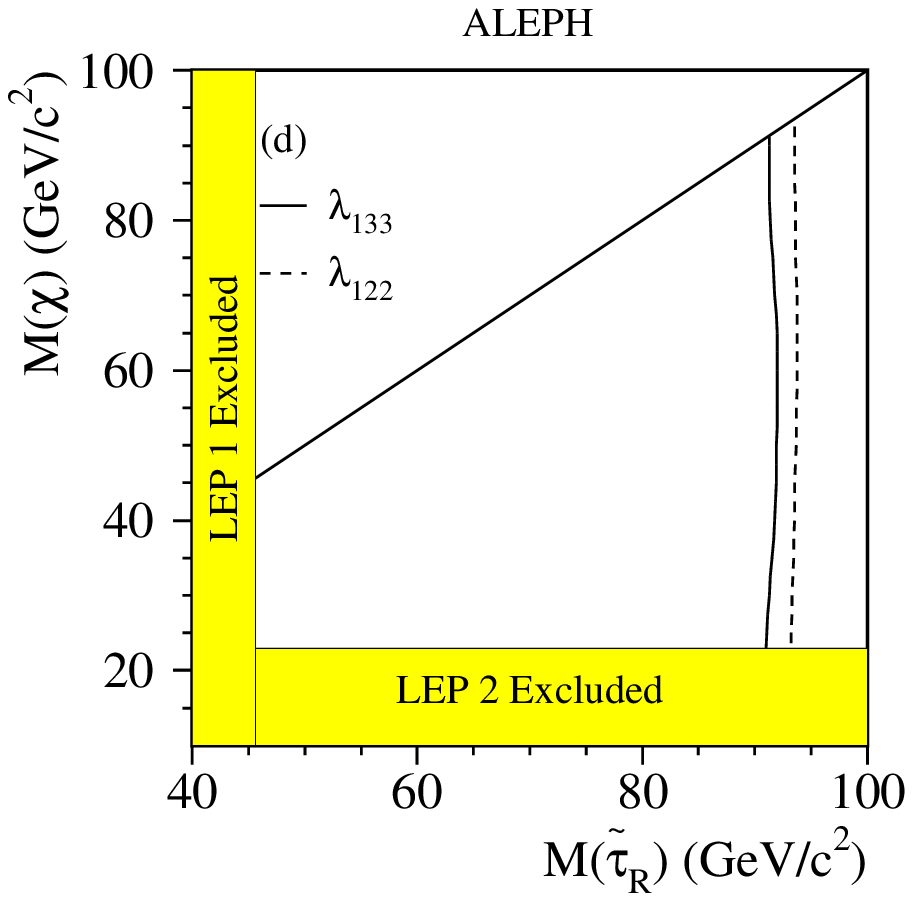}
}
\caption[.]{  \label{lle_sleptons}{a) \small The $95\%$ C.L. exclusion
cross sections for sleptons decaying directly via a dominant $\slle$
operator. The MSSM cross section for pair production of right-handed
selectrons and smuons are superimposed. Figures b), c) and d) show the
$95\%$ C.L. limits in the ($M_{\chi}$, $M_{\slR}$) plane for indirect
decays of selectrons, smuons and staus, respectively. The two choices of
$\lambda_{122}$ and $\lambda_{133}$ correspond to the best and worst
case exclusions, respectively. The selectron cross section is
evaluated at $\mu=-200~\gevcc$ and $\tanb=2$.}}
\end{center}
\end{figure}

 A right-handed slepton can decay directly via the $LL{\bar E}$
coupling to a lepton and anti-neutrino, hence the acoplanar lepton
selection is used. For a given choice of generation indices the decay
will produce two final states equally; for the coupling
$\lambda_{ijk}$ these decays are ${\tilde{l}_R}^k \ra
l^i{\bar{\nu}_{l^j}}$ or ${\bar{\nu}_{l^i}}l^j$. Excluded cross
sections are shown in Fig.~\ref{lle_sleptons}(a) for the different
mixtures of acoplanar lepton states. The MSSM production cross section
for right-handed smuon pairs and selectron pairs at $\mu =
-200~\gevcc$ and $\tanb=2$ are superimposed. The cross section limit
translates into a lower bound on the smuon (or stau) mass of
$M_{\smuR,\stauR}>81~\gevcc$ and $M_{\seR}>92~\gevcc$ ($\mu =
-200~\gevcc$, $\tanb=2$) for the direct decays and the worst case
coupling.

  Indirect decays of sleptons are selected using the ``Six Leptons $+
\emiss$'' selection. Limits corresponding to this case are shown in
Fig.~\ref{lle_sleptons}(b), (c), and (d).  Using the bound of
$M_{\chi}>23~\gevcc$ these limits can be interpreted as the mass
limits $M_{\seR}>93~\gevcc$ ($\mu=-200~\gevcc$, $\tanb=2$),
$M_{\smuR}>92~\gevcc$ and $M_{\stauR}>91~\gevcc$ for the worst case
coupling.

\subsection{Sneutrinos decaying via \boldmath $LL{\bar E}$ }

\begin{figure}
\begin{center}
\resizebox{\textwidth}{!}{
  \includegraphics{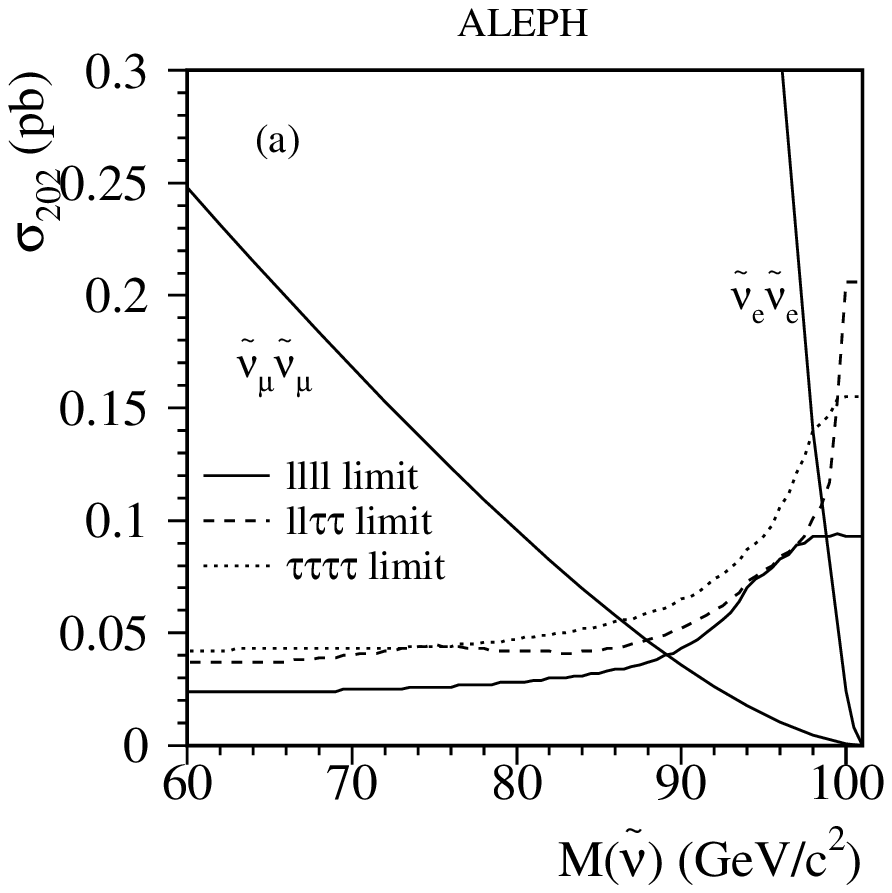}
  \includegraphics{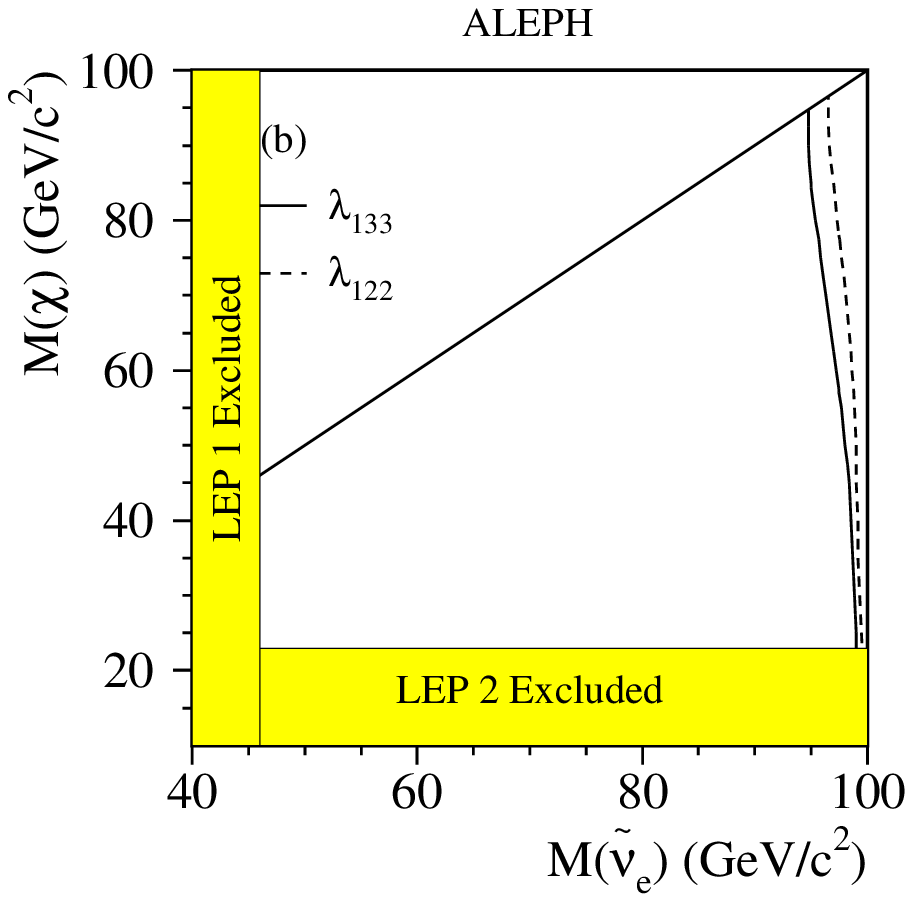}
}
\resizebox{0.5\textwidth}{!}{
  \includegraphics{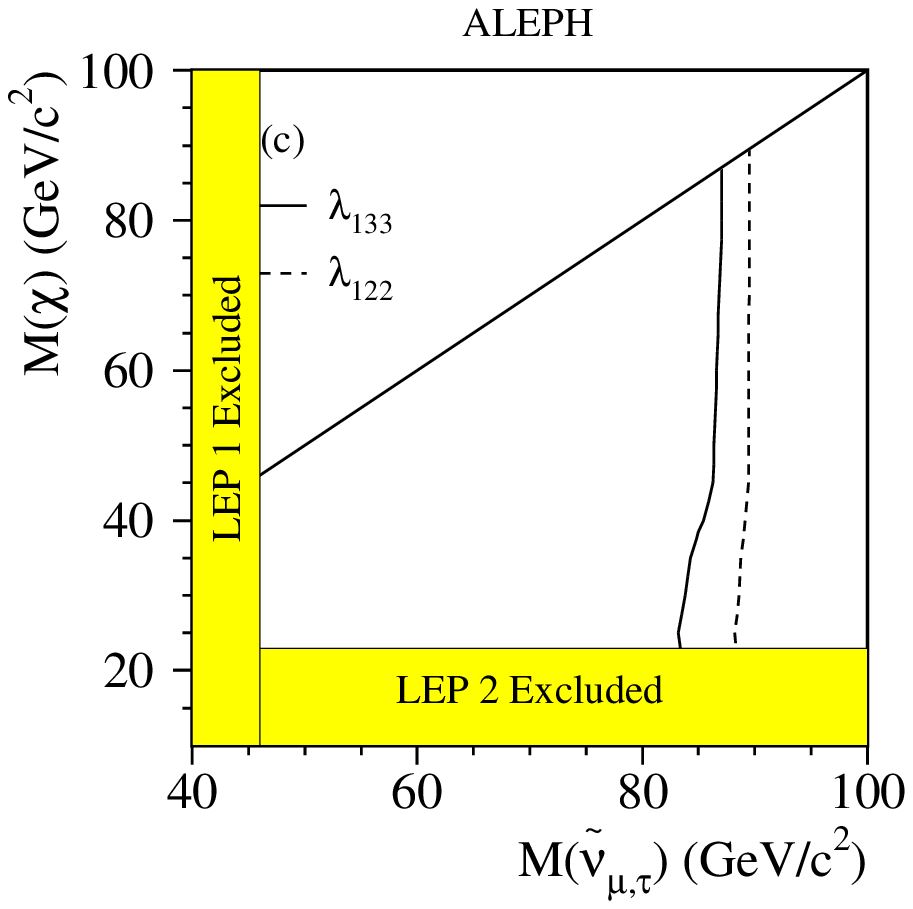}
}
\caption[.]{  \label{lle_sneus}{(a) \small The $95\%$ C.L. exclusion
cross sections for sneutrinos decaying directly via a dominant $\slle$
operator. The MSSM cross section for pair production of muon and
electron sneutrinos are superimposed; the tau sneutrinos have the same
cross section as the muon type. Figure (b) shows the $95\%$
C.L. limits in the ($M_{\chi}$, $M_{\snu}$) plane for $\snu_e$, and
figure (c) for both $\snu_{\mu}$ and $\snu_{\tau}$ indirect
decays. The two choices of $\lambda_{122}$ and $\lambda_{133}$
correspond to the best and worst case exclusions, respectively. The
electron sneutrino cross section is evaluated at $\mu=-200~\gevcc$ and
$\tanb=2$.}}
\end{center}
\end{figure}

  In pair production each sneutrino can decay directly into
pairs of charged leptons giving the final states $eeee$, $ee\mu\mu$,
$ee \tau \tau$, $\mu \mu \mu \mu$, $\mu \mu \tau \tau$ and $\tau \tau
\tau \tau$. The different final states correspond to
different choices of generation indices. The ``Four Lepton'' selection
was used to derive exclusion limits on the sneutrino pair production
cross section shown in Fig.~\ref{lle_sneus}(a).  These limits
translate into a lower bound on the electron sneutrino mass of
$M_{\snue}>98~\gevcc$ ($\mu=-200~\gevcc$, $\tanb=2$) and the muon
sneutrino mass of $M_{\snumu}>86~\gevcc$ for direct decays and
the worst case coupling.

  Indirect decays of sneutrinos are selected using the ``Four Leptons
$+ \emiss$'' selection. The limits in the ($M_{\chi}$, $M_{\snu}$)
plane corresponding to this case are shown in Fig.~\ref{lle_sneus}(b)
and (c).  Using the bound $M_{\chi}>23~\gevcc$
this limit can be interpreted as $M_{\snu_{\mu,\tau}} > 83~\gevcc$ and
$M_{\snue} > 94~\gevcc$ for the worst case coupling, where the cross
section for the electron sneutrino is evaluated at $\mu=-200~\gevcc$
and $\tanb=2$.

\section {\label{lqd} Decays via a dominant \boldmath $LQ{\bar D}$ Coupling}

For a dominant $LQ{\bar D}$ operator the event topologies are mainly
characterised by large hadronic activity, possibly with some leptons
and/or missing energy. In the simplest case the topology consists of
four-jet final states, and in more complicated scenarios of
multi-jet and multi-lepton and/or multi-neutrino states. A summary 
of the results of the various selections is given in Table~\ref{topslqd}.

The acoplanar jet selection (AJ-H) and the four jets and missing
energy selection (4JH) developed for R-parity conserving SUSY searches
are used~\cite{rpc183}.  The ``MultiJets + Leptons'' and the
``2J+2$\tau$'' selections are updated from Ref.~\cite{183paper} and
reoptimised for the higher centre-of-mass energy as described
below. The ``Jets-HM'' selection is a retuned version of the
``MultiJets + Leptons'' subselection I for a high visible mass
system. The ``4 Jets + $2\tau$'' is unchanged from
Ref.~\cite{LQDpaper}. Two new selections for five jets and one
isolated lepton and four jets and two isolated leptons, ``5 Jets + 1
Iso.~$l$'' and ``4 Jets + 2 Iso.~$l$'', are described below.

\begin{table}
\caption[.]{ \small 
The observed numbers of events in the data and the Standard Model  
background expectations for the $\slqd$ selections.}
\label{topslqd}
\setlength{\tabcolsep}{0.12cm}
\begin{center}
\begin{tabular}{|c|cc|cc|cc|cc|cc||cc|}
\hline
\multicolumn{1}{|r|}{$\roots$ (GeV)} &
\multicolumn{2}{c|}{188.6} & \multicolumn{2}{c|}{191.6} &
\multicolumn{2}{c|}{195.5} & \multicolumn{2}{c|}{199.5} &
\multicolumn{2}{c||}{201.6} &\multicolumn{2}{c|}{All} \\
Selection & 
Data & SM & Data & SM &  Data & SM & Data & SM & Data & SM & Data & SM \\
\hline
\hline
MultiJets +&  
13 &  11 &  1 &   1.8 &  6 &   4.7 &  3 &   5.2 &  5 &   2.6 &  28 &
26 \\
Leptons   &&&&&&&&&&&& \\
\hline

Jets-HM &
11 &   8.7 &  1 &   1.3 &  3 &   3.1 &  1 &   2.8 &  3 &   1.3 &  19 &
17 \\
  &&&&&&&&&&&& \\
\hline

4 Jets + 2$\tau$ &
10 &  13 &  2 &   2.0 &  1 &   5.1 &  7 &   5.3 &  5 &   2.7 &  25 &
28 \\
&&&&&&&&&&&& \\
\hline

Four-Jets &
684 & 754 &143 & 127 &322 & 351 &336 & 370 &147 & 179 &1632
& 1780 \\
&&&&&&&&&&&& \\
\hline

2 Jets + 2$\tau$ &
10 &  11 &  1 &   1.9 &  5 &   5.8 &  6 &   5.2 &  3 &   2.5 &  25 &
26 \\
&&&&&&&&&&&& \\
\hline

AJ-H &
10 &  12 &  2 &   2.2 &  9 &   6.8 &  7 &   8.6 & 10 &   4.6 &  38 &
34 \\
&&&&&&&&&&&& \\
\hline

4JH &
 5 &   7.8 &  1 &   1.3 &  3 &   3.7 &  0 &   3.9 &  4 &   2.0 &  13 &
19 \\
&&&&&&&&&&&& \\
\hline

5 Jets + &
4 &  4.2 &  1 &  0.7 &  1 &  1.9 &  0 &  1.9 &  0 &  0.9 &   6 &
9.6 \\
1 Iso.~$l$  &&&&&&&&&&&& \\
\hline

4 Jets + &
0 &     3.1 &0  &  0.5 &    1  &  1.3&     3 &   1.5 &    1  &  0.7 &
5 & 7.1\\
2 Iso.~$l$  &&&&&&&&&&&& \\
\hline
\end{tabular}
\end{center}
\end{table}

\subsection{Charginos and neutralinos decaying via \boldmath $LQ{\bar D}$ }

Three subselections were developed to select the chargino
indirect topologies~\cite{183paper}; some cuts have been reoptimised
for the higher centre-of-mass energy. Subselection~I is designed to
select final states based on hadronic activity, e.g. $\chi^+\chi^-
\ra qqqq\chi\chi$; subselection~II is designed for decays such as
$\chi^+\chi^- \ra l\nu qq\chi\chi$ where the leptonic energy is
larger, and subselection~III is designed to select the decays
$\chi^+\chi^- \ra l\nu l\nu\chi\chi$. The combination of the three
subselections is defined as the ``Multi-jets plus Leptons'' selection.
The complete set of cuts is shown in Table~\ref{mjlcuts}.

Interpreting these results in the framework of the MSSM, 95\%
C.L. exclusion limits are derived in the ($\mu$,$M_2$) plane and shown
in Fig.~\ref{mum2lims}(b) for large scalar masses
($m_0=500~\gevcc$). The corresponding lower limit on the mass of the
lightest chargino is essentially at the kinematic limit for pair
production.

The searches for the lightest and second lightest
neutralino do not extend the excluded region in the
($\mu$,$M_2$) plane beyond that achieved with the chargino 
search alone.

\begin{table}[t]
\caption{\small \label{mjlcuts} The list of cuts for the
``Multi-jets plus Leptons'' selection used to select indirect chargino
decays via the $\slqd$ operator. The ``Jets-HM'' selection is also
listed; this is used to select intermediate $\Delta{M}$ indirect
squark decays with high visible mass. Primed event variables are
calculated from physical quantities excluding identified
leptons. The event variables are defined in Ref.~\cite{183paper}.}
\begin{center}
\begin{tabular}{|c|c|c||c|} \hline
 Subselection~I & Subselection~II & Subselection~III &Jets-HM \\
\hline \hline
\multicolumn{3}{|c||}{$\nch \geq 10$}&$\nch \geq 30$ \\
\multicolumn{3}{|c||}{$\mvis > 45~\gevcc$} &
$\mvis > 100~\gevcc$\\
\multicolumn{3}{|c||}{$\Theta_\mathrm{miss} > 30^\circ$} &
$\Theta_\mathrm{miss} > 30^\circ$ \\
\hline
$\mvis'>50\%\sqrt{s}$ & $\mvis'<50\%\sqrt{s}$ & $\mvis'<65~\gevcc$ &
$\mvis'>50\%\sqrt{s}$ \\
$T < 0.9$ & $T <0.74$ & $T<0.8$ &
$T < 0.85$ \\
 & $y_4' > 0.0047$ & $y_4'>0.001$ &\\
$y_5>0.003$ & & &
$y_5>0.003$ \\
$y_6>0.002$ &  & $y_6>0.00035$ &
$y_6>0.002$ \\
$E_T>80~\gev$ & \raisebox{2.5ex}[-2.5ex]
{$\left (\begin{array}{c}
\Phi_\mathrm{aco}'<145^\circ\\
\mathrm{or} \\
y_6 > 0.002
\end{array} \right )$}& &$E_T>80~\gev$ \\
\hline
$E^\mathrm{em}_\mathrm{jet} < 90\%E_\mathrm{jet}$ & $\elep\ < 40~\gev$ &  &
$E^\mathrm{em}_\mathrm{jet} < 85\%E_\mathrm{jet}$ \\
$E^\mathrm{iso}_\mathrm{10} < 5~\gev$  & $\ehad\ < 2.5 \elep$ & $\ehad\ < 47\%\elep$&
$E^\mathrm{iso}_\mathrm{10} < 5~\gev$ \\
\hline
$[0.55(\mvis'-120)+{}$&\multicolumn{2}{c||}{$\chiWW>$~3.8}&
$[0.55(\mvis'-120)+{}$\\
$\Phi_\mathrm{aco}']<180^\circ$ &\multicolumn{2}{c||}{}&
$\Phi_\mathrm{aco}']<190^\circ$ \\
\hline
\end{tabular}
\end{center}
\end{table}

\subsection{Squarks decaying via \boldmath $LQ{\bar D}$ }

\begin{figure}
\begin{center}
\resizebox{\textwidth}{!}{
  \includegraphics{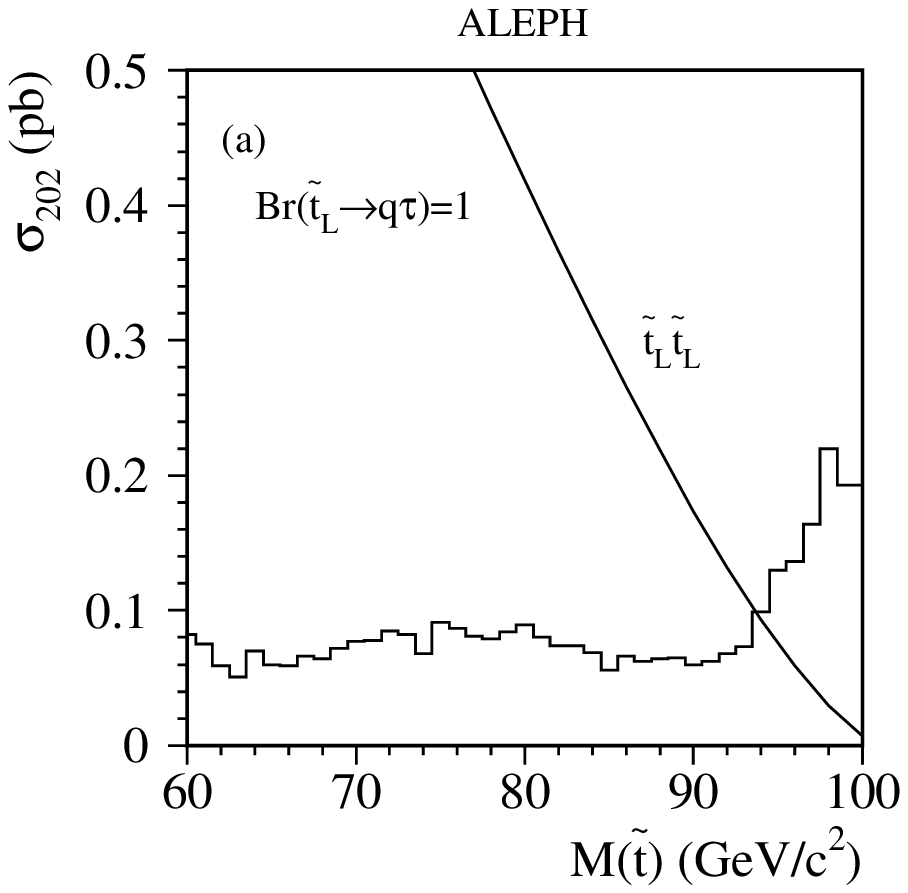}
  \includegraphics{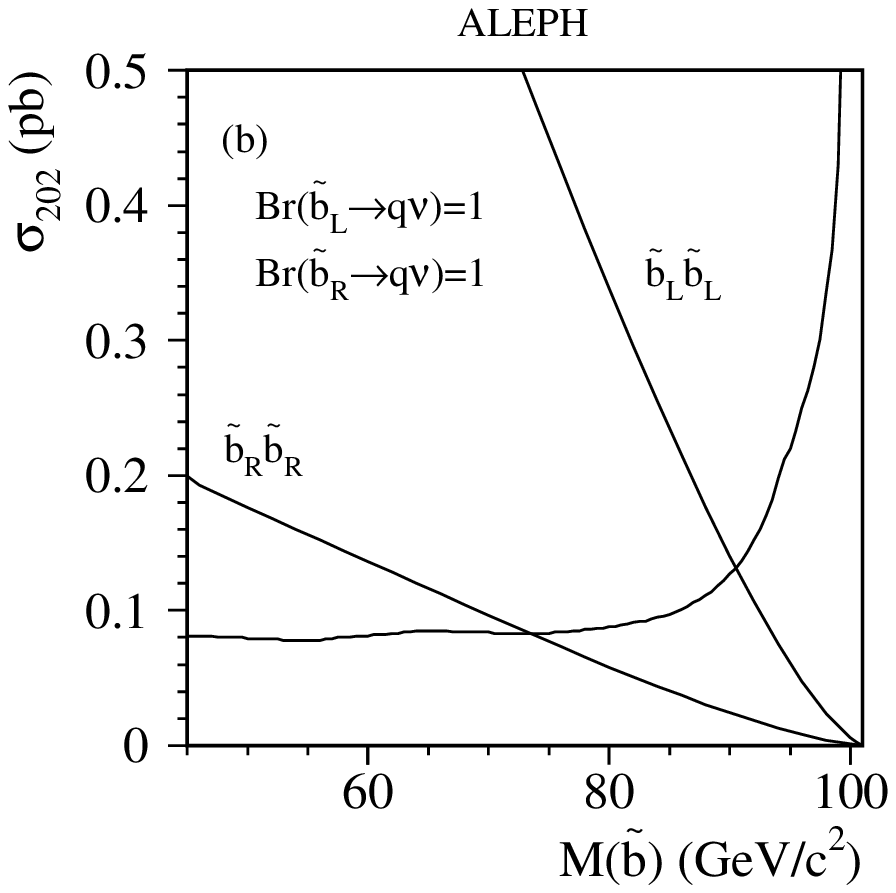}
}
\resizebox{0.5\textwidth}{!}{
  \includegraphics{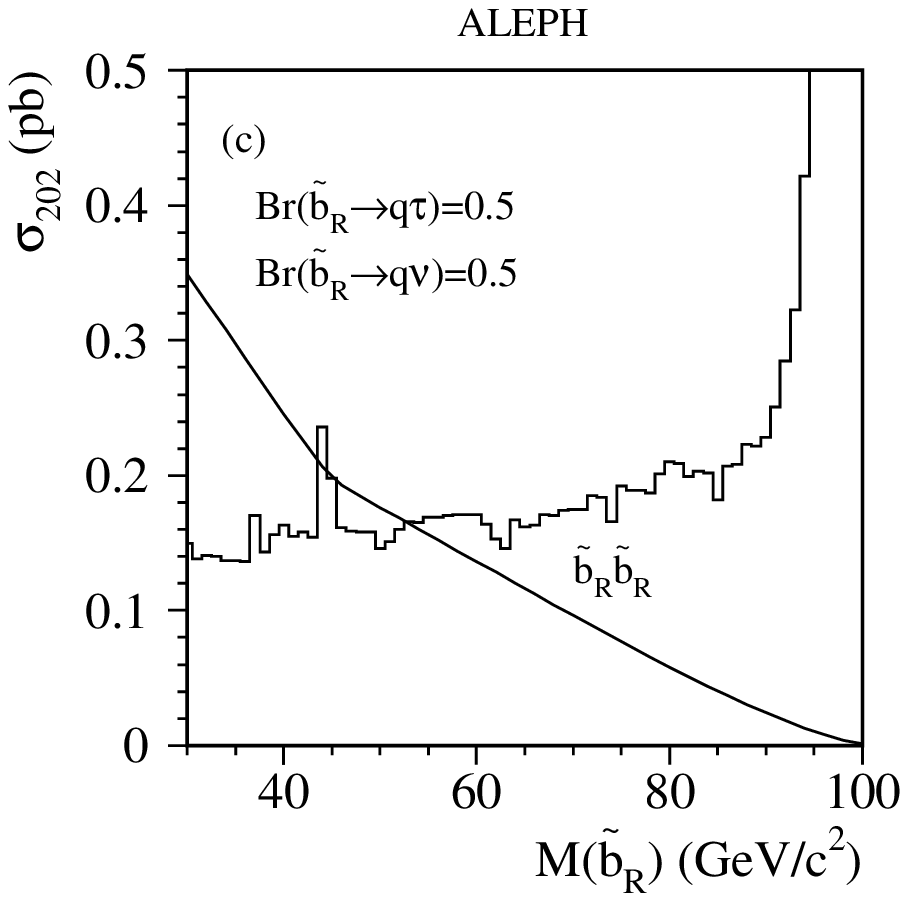}
}
\caption[.]{  \label{sq.direct}{\small 
The $95\%$ C.L. excluded cross sections for the production of squarks
decaying directly via a dominant $\slqd$ operator: (a) $\stL$
$(\lambda'_{33k})$, (b) $\sbL$ $(\lambda'_{i3k})$ or $\sbR$
$(\lambda'_{i33})$, and (c) $\sbR$ $(\lambda'_{3j3})$. The MSSM cross
sections are superimposed.}}
\end{center}
\end{figure}

A squark can decay directly to a quark and either a lepton or a
neutrino leading to topologies with acoplanar jets and up to two
leptons. Couplings with electrons or muons in the final state are not
considered as existing limits from the Tevatron~\cite{TeVLQ} exclude
the possibility of seeing such a signal at LEP.  To select $\sq \ra \q
\tau$ and $\sq \ra \q\nu$, the ``2J+2$\tau$'' and the ``AJ-H''
selections are used.  The ``2J+2$\tau$'' selection is unchanged from
Ref.~\cite{183paper} except that the sliding mass window is now
$10~\gevcc$ wide centred on the squark mass. Examples of limits for
squark production are shown in Fig.~\ref{sq.direct}. In particular,
for a dominant $\lambda'_{33k}$ coupling, which implies Br$(\stL
\rightarrow \mathrm{q} \tau)=100\%$, a lower limit of $93~\gevcc$ is
obtained for~$M_{\stL}$.

\begin{figure}
\begin{center}
\resizebox{\textwidth}{!}{
  \includegraphics{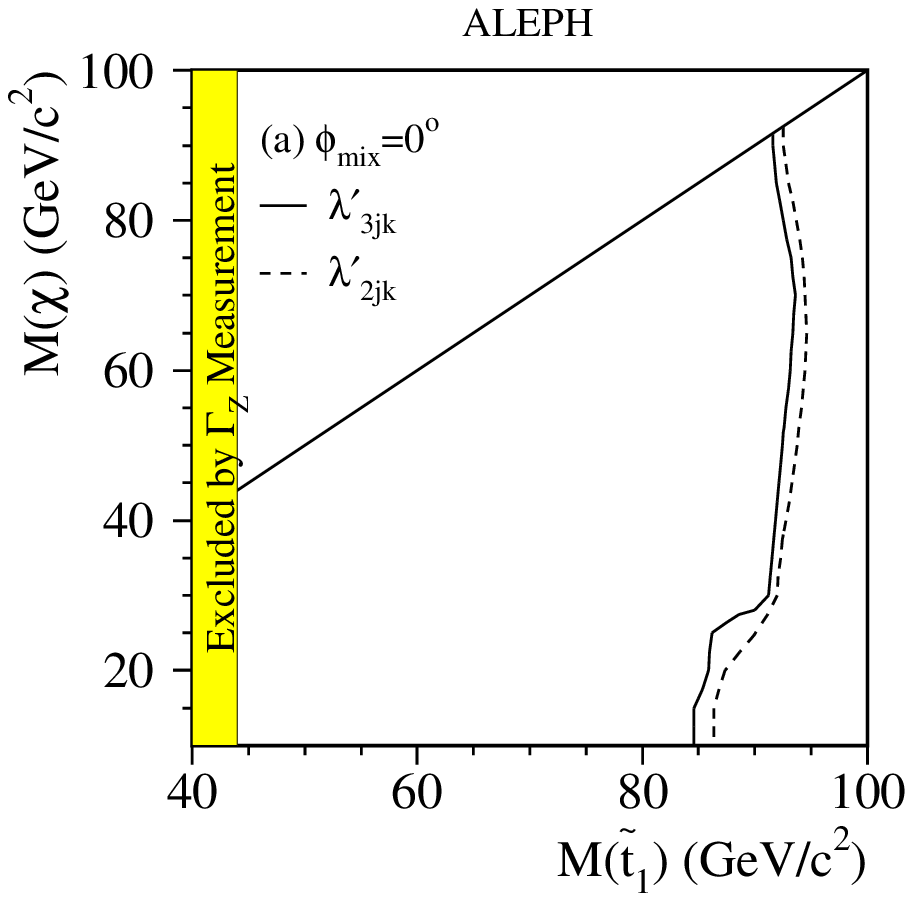}
  \includegraphics{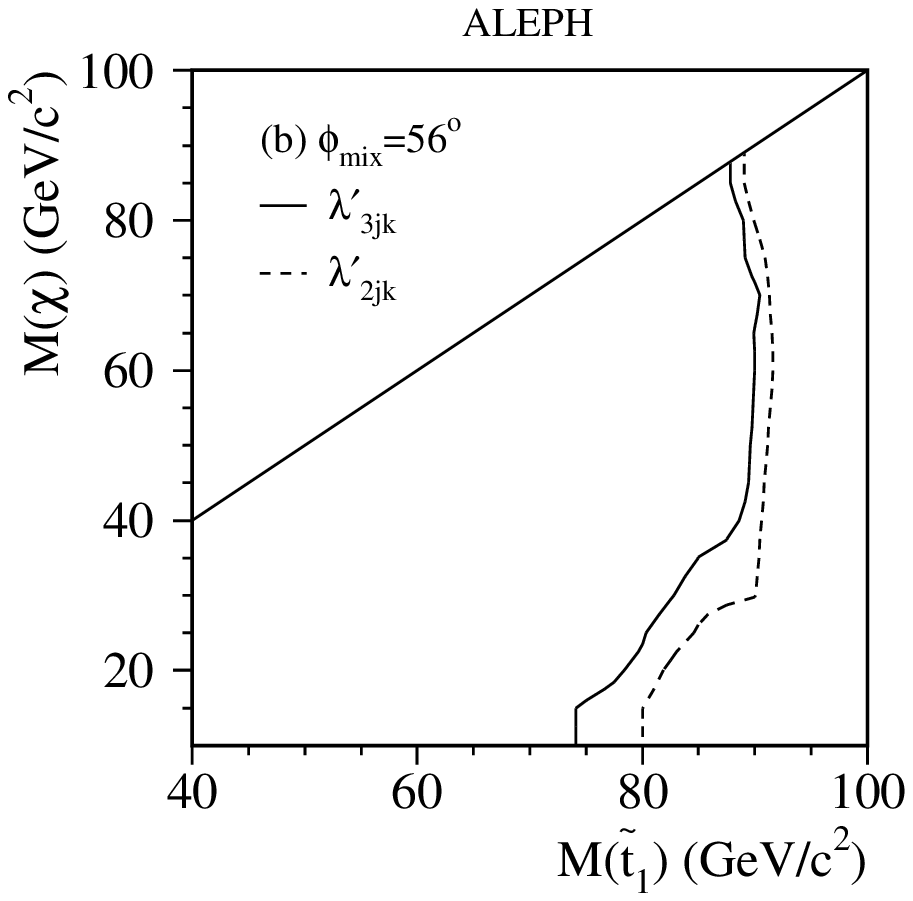}
}
\resizebox{\textwidth}{!}{
  \includegraphics{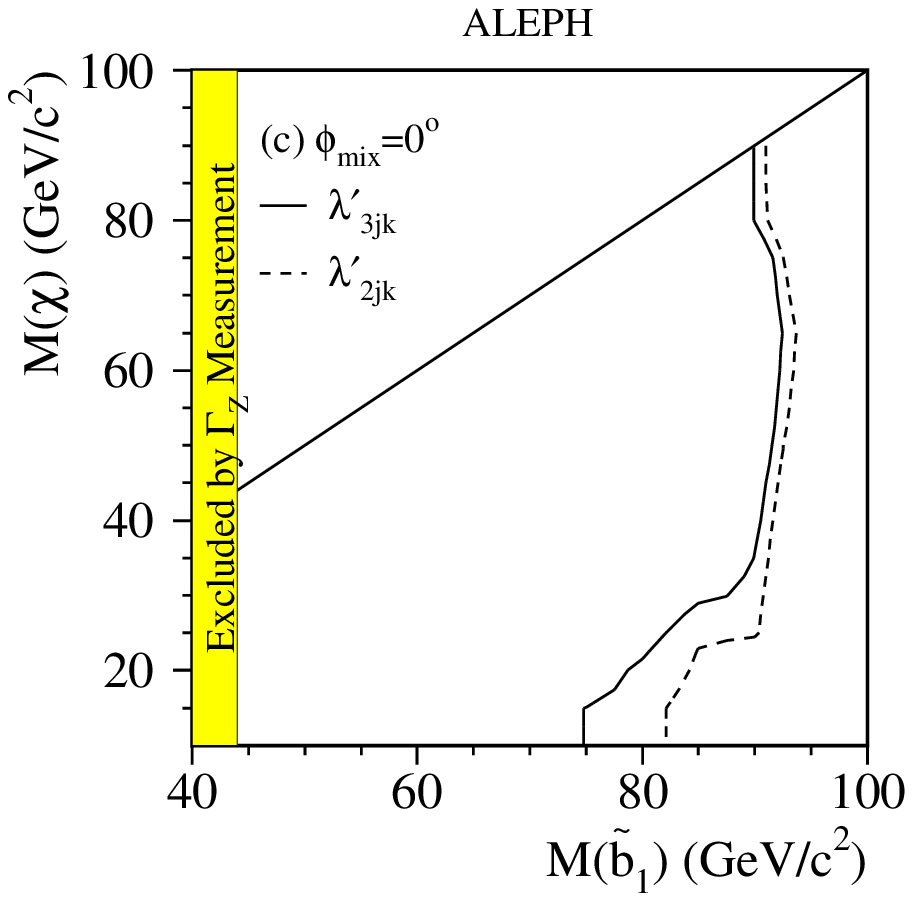}
  \includegraphics{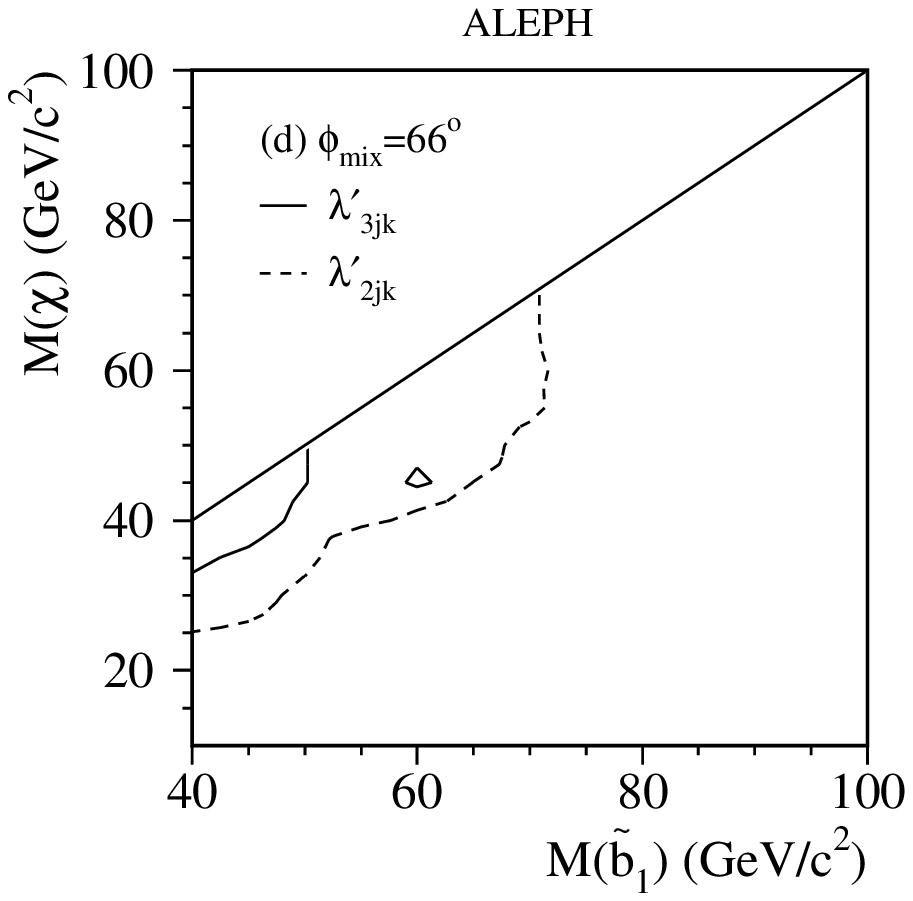}
}
\caption[.]{\label{fig:squark_lqd_ind.eps} {\small The $95\%$ C.L. limits in the
($M_{\chi}$, $M_{\stone}$) and ($M_{\chi}$, $M_{\sbone}$) planes for
indirect $\slqd$ decays via the $\lambda'_{211}$ and 
$\lambda'_{311}$ couplings, 
for no squark mixing ($\phimix=0^\circ$)
and for $\phimix=56^\circ,66^\circ$
for stops and sbottoms, respectively.
}}
\end{center}
\end{figure}

 Indirect decays of squarks via the $LQ{\bar D}$ operator will give
six jets and up to two charged leptons. These are selected by the ``4
Jets + 2$\tau$'' selection if $M_{\chi}\leq20~\gevcc$, either the ``5
Jets + 1 Iso.~$l$'' or ``Multi-jets plus Leptons'' if
$(M_{\sq}-M_{\chi})\leq15~\gevcc$, otherwise the ``Jets-HM'' selection
is used. The cuts used for the ``5 Jets + 1 Iso.~$l$'' selection are
listed in Table~\ref{table:4J_12IsoL}. Limits for the optimistic case
of left-handed squarks are shown in
Fig.~\ref{fig:squark_lqd_ind.eps}. The following limits for $\stL$ and
$\sbL$ are derived: $M_{\stL}>84~\gevcc$ and $M_{\sbL}>74~\gevcc$.

\begin{table}
\caption[FJ+1isolep]{\small \label{table:4J_12IsoL}
The list of cuts for the ``5 Jets + 1 Isolated Leptons'' and ``4 Jets
+ 2 Isolated Leptons'' selections, as used for the indirect
squark and slepton searches decaying via an $\slqd$
operator. $M''_{vis}$ is the visible mass after the two leading
leptons are removed. Other variables are
defined in Ref.~\cite{183paper}.}
\begin{center}
\begin{tabular}{|c|c|}
\hline
5 Jets + 1 Iso.~$l$ & 4 Jets + 2 Iso.~$l$\\
\hline
\hline
$\nch > 9$   & $\nch > 9$ \\
$\evis < 95\% \roots$ & $\evis > 65\% \roots$ \\
$N_\mathrm{lep} \geq 1$ & 
  $N_\mathrm{lep} \geq 2$ (same flavour e or $\mu$) \\
$\ptmiss>5~\gevc$ & \\
$\Theta_\mathrm{miss} > 18^\circ$ &  \\
\hline
$E_{l_1}>10~\gev$ & $E_{l_1}>5~\gev$, $E_{l_2}>5~\gev$ \\
$E_{l_1}^{iso}<5~\gev$ & 
  $E_{l_1}^\mathrm{iso}<5~\gev$, $E_{l_2}^\mathrm{iso}<5~\gev$\\
\hline
$y_5>0.003$ & $y_5>0.001$\\
$y_6>0.001$ & $y_6>0.0005$\\
\hline
$\min\left\{{\left|{M_\mathrm{vis}-91.2}\right|,
\left|{M''_\mathrm{vis}-91.2}\right|}\right\}>3~\gevcc$ & 
  $|M''_\mathrm{vis}-91.2|>5~\gevcc$ \\
\hline
\multicolumn{2}{|c|}{$\chiWW>$~3.8}\\
\hline
\end{tabular}
\end{center}
\end{table}

\subsection{Sleptons and sneutrinos decaying via \boldmath $LQ{\bar D}$ }
\label{lqd.fourjets}
\begin{figure}
\begin{center}
\resizebox{\textwidth}{!}{
    \includegraphics{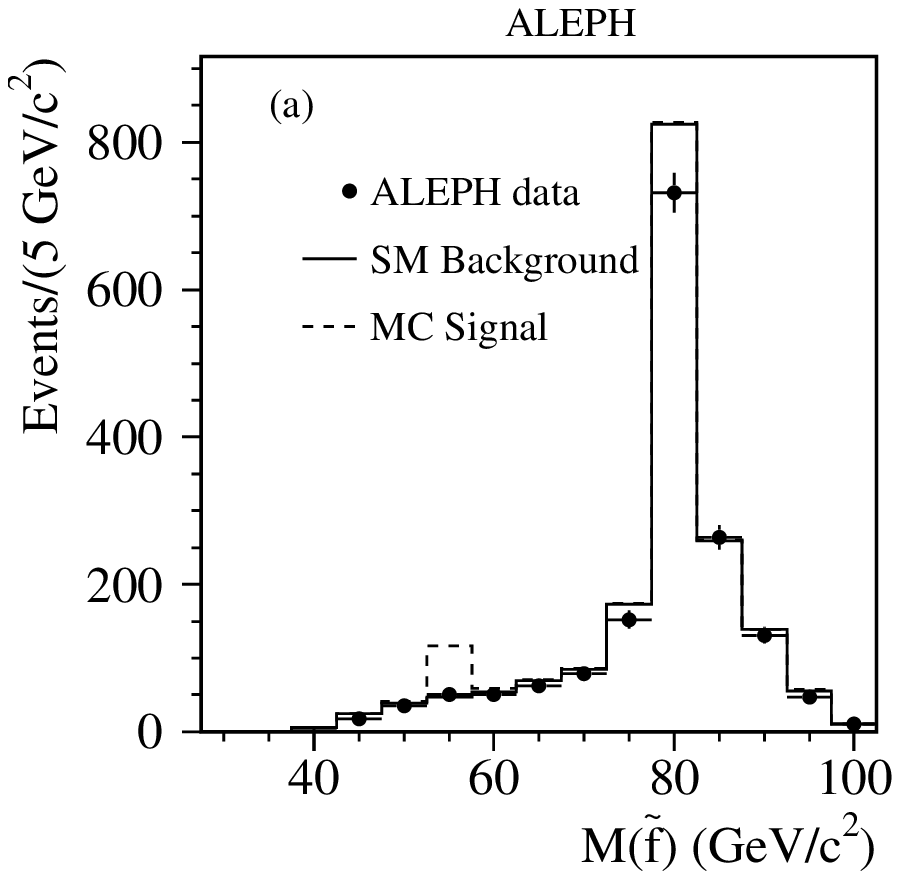}
    \includegraphics{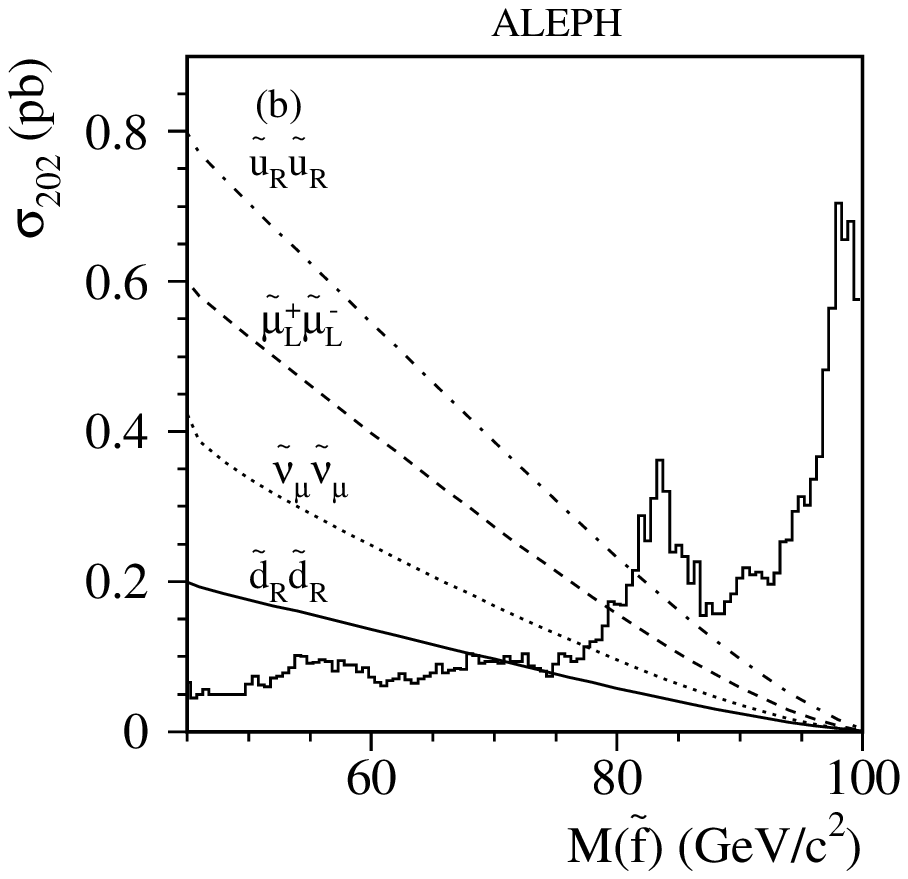}}
\caption[.]{  \label{fig:fourjets} \small
(a): The distribution of the reconstructed invariant masses for jet
pairs after forcing the event into four jets. The points are the data
(from 188.6 GeV to 201.6 GeV), the solid histogram is the Monte Carlo
predicted background. The dashed histogram is the signal for the
process $\epem\ra\smuL^+\smuL^-$ with the smuons decaying directly via
the $\slqd$ coupling; here $M_{\smuL}=55~\gevcc$ and the histogram is
normalised to the expected cross section for this process
($\sigma=0.46~\mathrm{pb}$ at $\roots=202~\gev$).  (b): The $95\%$
C.L. excluded cross sections for sleptons (via $\slqd$), sneutrinos
(via $\slqd$) and squarks (via $\sudd$) decaying directly to four
jets. The MSSM cross sections for pair production of muon sneutrinos,
left-handed smuons and right-handed squarks are superimposed.  }
\end{center}
\end{figure}

  Direct decays of sleptons and sneutrinos via the $LQ{\bar D}$
operator lead to four jet final states. The ``Four-Jets'' selection
from Ref.~\cite{183paper} is applied.  The distributions of the di-jet
masses for data and Monte Carlo are shown in
Fig.~\ref{fig:fourjets}(a).  Fewer events are observed around the W
mass peak region than expected.  Limits are derived by sliding a mass
window of $5~\gevcc$ across the di-jet mass distribution. The results
are shown in Fig.~\ref{fig:fourjets}(b) and imply
$M_{\snumu}>77~\gevcc$ and $M_{\smuL}>79~\gevcc$.

\begin{figure}
\centering
\resizebox{\textwidth}{!}{
  \includegraphics{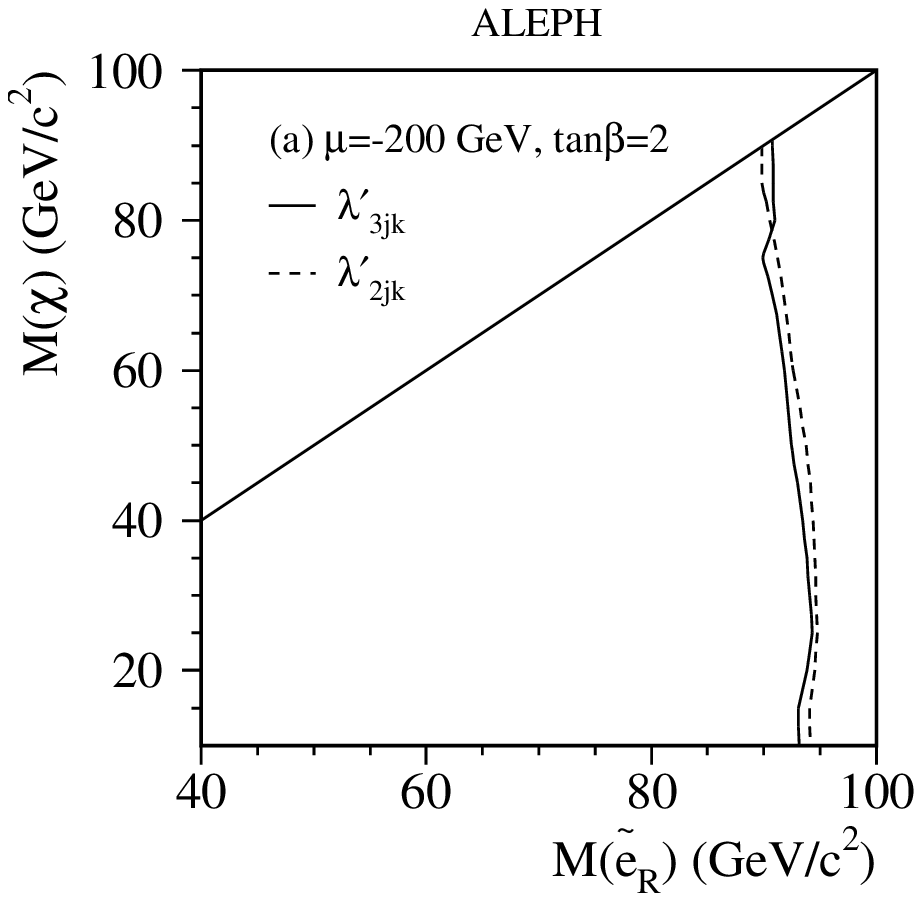}
  \includegraphics{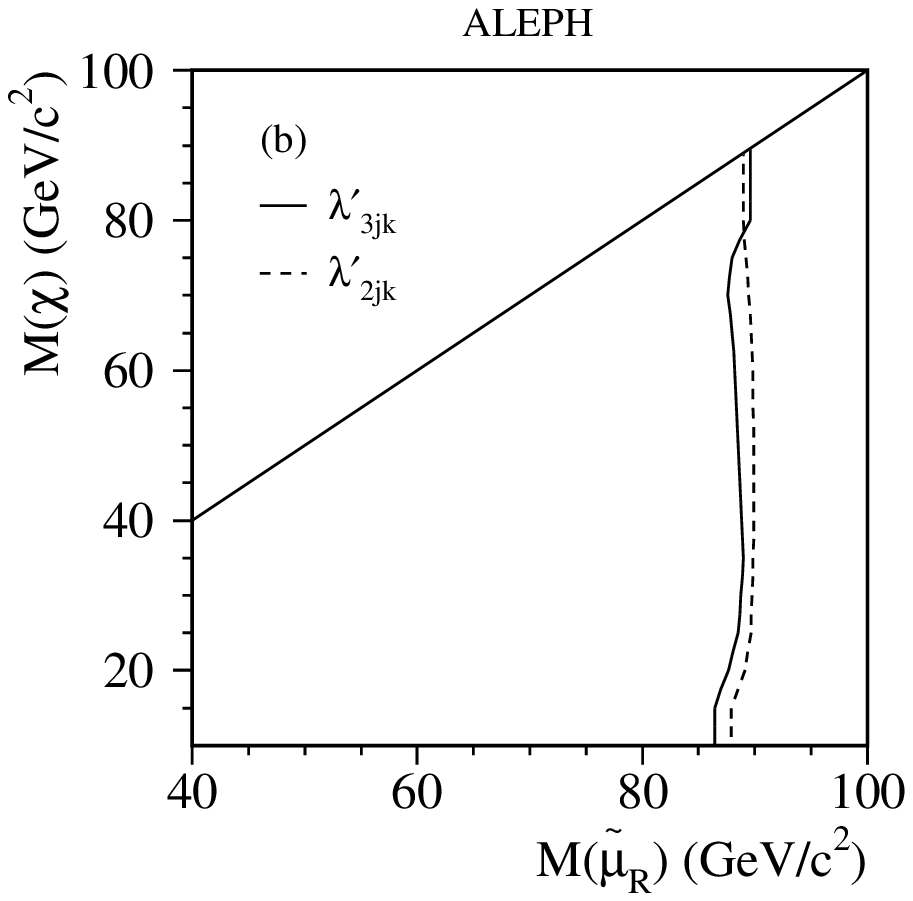}
}
\resizebox{0.5\textwidth}{!}{
  \includegraphics{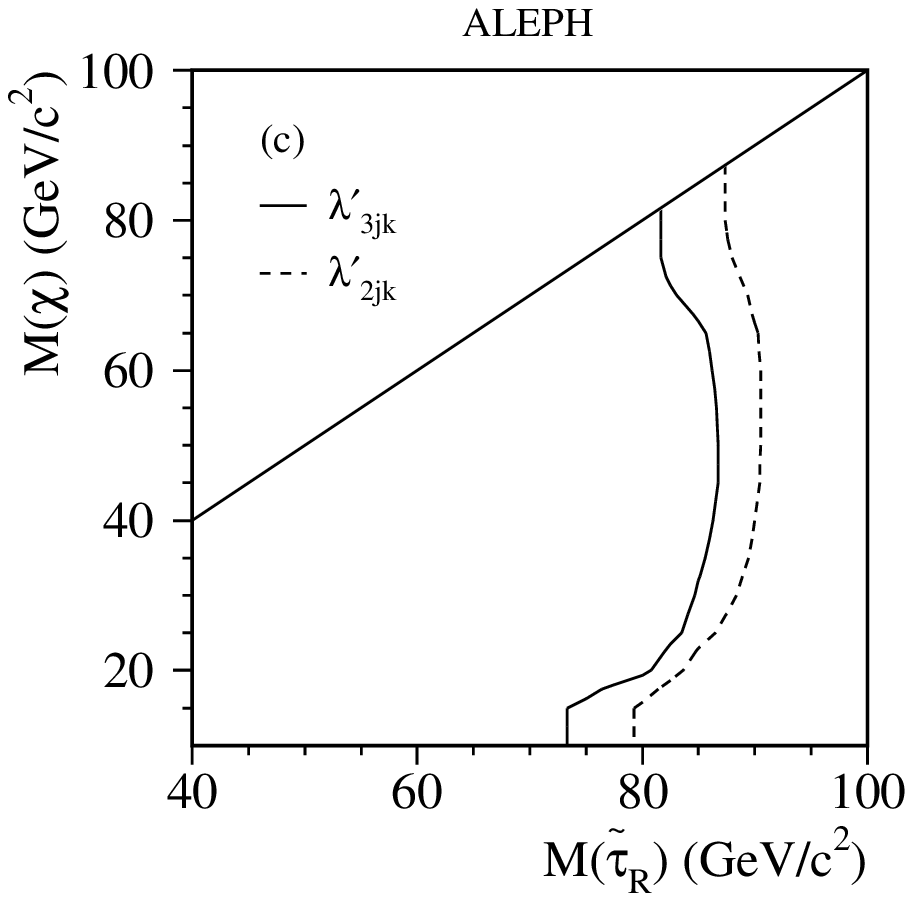}
}
\caption[.]{  \label{fig:slep_LQD_ind} \small
The $95\%$ C.L. limits in the ($M_{\chi}$, $M_{\slR}$)
plane for selectrons, smuons and staus decaying indirectly via a
dominant $\slqd$ operator. The two choices of $\lambda'_{2jk}$ and
$\lambda'_{3jk}$ correspond to the best and worst case exclusions,
respectively. The selectron cross section is evaluated 
at $\mu=-200~\gevcc$ and $\tanb=2$. The limit from the $\Gamma_\mathrm{Z}$
measurements excludes $M_{\slep}<38~\gevcc$.  }
\end{figure}

\begin{figure}
\centering
\resizebox{\textwidth}{!}{
  \includegraphics{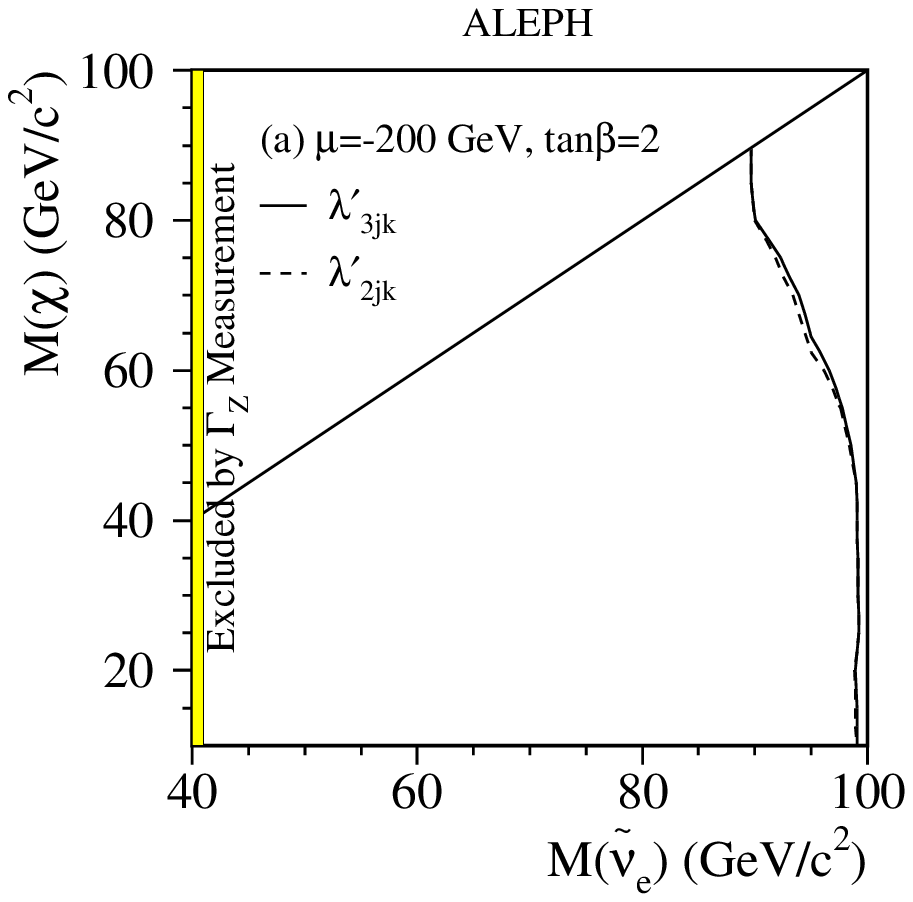}
  \includegraphics{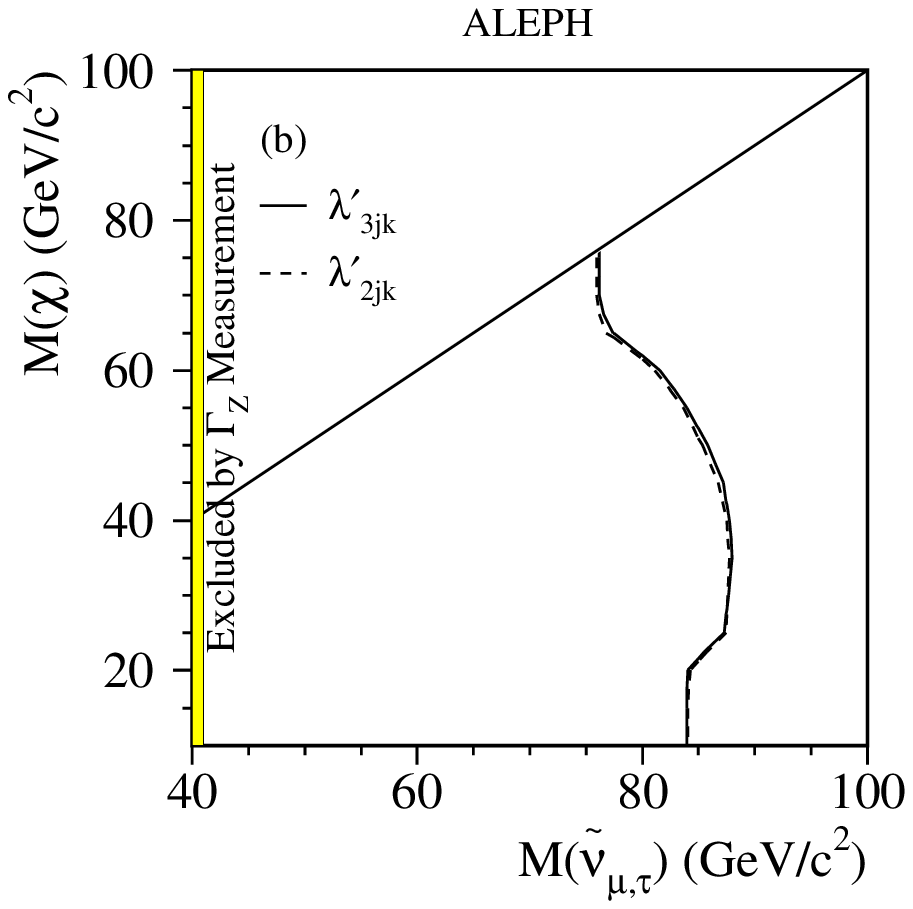}
}
\caption[.]{  \label{fig:snu_LQD_ind} \small
 The $95\%$ C.L. limits in the ($M_{\chi}$, $M_{\snu}$) plane for
electron and muon or tau sneutrinos decaying indirectly via a dominant
$\slqd$ operator. The two choices of $\lambda'_{2jk}$ and
$\lambda'_{3jk}$ correspond to the best and worst case exclusions,
respectively. The electron sneutrino cross section is evaluated at
$\mu=-200~\gevcc$ and $\tanb=2$.}
\end{figure}

Indirect decays of the sleptons via the $LQ{\bar D}$ operator will
give two, three or four leptons and four jets in the final state; two
leptons will be of the same flavour as the initial sleptons.  The
indirect decays of sneutrinos will give a final state with four jets,
up to two leptons and missing energy. For selectrons and smuons the
``4 Jets + 2 Iso.~$l$'' selection is used except for the
special case of $\lambda'_{3jk}\neq0$ and
$(M_{\slR}-M_\chi)<10~\gevcc$ where the ``4 Jets + 2$\tau$'' selection
is used. The cuts used in the ``4 Jets + 2 Iso.~$l$'' are listed in
Table~\ref{table:4J_12IsoL}. Indirect stau decays are selected
with the ``5 Jets + 1 Iso.~$l$'' selection if $M_\chi>20~\gevcc$ and
either $\lambda'_{2jk}\neq0$ or $\lambda'_{1jk}\neq0$, otherwise the
inclusive combination of the ``5 Jets + 1 Iso.~$l$'' and the ``Leptons
and Hadrons'' selections is used. The sneutrinos are selected with the
``4JH'' selection for $M_\chi>20~\gevcc$ and ``AJ-H'' (acoplanar jets)
otherwise. Limits for these decays are shown in
Fig.~\ref{fig:slep_LQD_ind} and Fig.~\ref{fig:snu_LQD_ind}. The limits
are $M_{\seR}>89~\gevcc$, $M_{\smuR}>86~\gevcc$,
$M_{\stauR}>73~\gevcc$, $M_{\snue}>89~\gevcc$ and
$M_{\snumu}>75~\gevcc$ for the worst case couplings; the selectron
and electron sneutrino cross sections are evaluated at
$\mu=-200~\gevcc$ and $\tanb=2$.

\section {\label{udd} Decays via a dominant \boldmath 
${\bar U}{\bar D}{\bar D}$ coupling}

For a dominant $\sudd$ operator the final states are characterised 
by topologies having many hadronic jets, possibly associated with 
leptons and missing energy. A number of selections are used:
``Four Jets'', ``Four Jets-Broad'', ``Many Jets'', ``Many Jets + Leptons'',
``Four Jets + 2 Leptons'', ``Many Jets + 2 Leptons'',
``Four Jets + $\emiss$'' and ``Many Jets + $\emiss$'', all introduced
in Ref.~\cite{183paper}. 
The latter two selections, which require missing energy,   
have been slightly modified for the higher energy and are summarised
in Table \ref{UDDcuts_sneu}. These selections rely mainly
on two characteristics of the events: mass reconstruction of the pair
produced sparticles and/or the presence of many jets in the event.
Table \ref{topsUDD} gives a list of all the selections 
and the numbers of observed and expected events. 

\begin{table}[t]
\caption{\small \label{UDDcuts_sneu}
The list of cuts for the ``Four Jets + $\emiss$''
and ``Many Jets + $\emiss$'' selections, as used for the 
indirect sneutrino searches via a $\sudd$ operator.}
\begin{center}
\begin{tabular}{|c|c|} \hline
 Four Jets + $\emiss$ & Many Jets + $\emiss$ \\
\hline \hline
\multicolumn{2}{|c|}{$\nch\ > 8$}\\
\multicolumn{2}{|c|}{$|\pzmiss|/\pmiss<0.95$} \\
\multicolumn{2}{|c|}{$E^\mathrm{em}_\mathrm{jet} < 95\%E_\mathrm{jet}$, $E_T>60~\gev$, $\elep\ < 15~\gev$} \\
\hline
$0.25<E_\mathrm{vis}/\sqrt s<0.75$  &  $0.5<E_\mathrm{vis}/\sqrt s<0.95$          \\
$\Delta\phi_T < 170^\circ$  &                                              \\
$0.5<T<0.97$                 &  $0.6<T<0.97$                               \\
\hline
$y_4 >0.001$                &  $y_4 >0.005$                                \\
$y_6>0.0003$                &  $y_6>0.002$ ($0.005$ if $M_{\chi}>60~\gev$) \\
\hline
$|M_{12-34}|<10~\gevcc$    & if $M_{\chi}<60~\gevcc$: $|M_{12-34}|<10~\gevcc$\\
$|M_{12+34}-2M_{\chi}|<M_{\chi}/3$ & if $M_{\chi}<60~\gevcc$: $|M_{12+34}-2M_{\chi}|<M_{\chi}/3$ \\
\hline
\end{tabular}
\end{center}
\end{table}

\begin{table}
\caption[.]{ \small 
The observed numbers of events in the data and the corresponding Standard Model
background expectations for the $\sudd$ selections, 
quoted with any sliding mass cuts removed.}
\label{topsUDD}
\setlength{\tabcolsep}{0.12cm}
\begin{center}
\begin{tabular}{|c|cc|cc|cc|cc|cc||cc|}
\hline
\multicolumn{1}{|r|}{$\roots$ (GeV)} &
\multicolumn{2}{c|}{188.6} & \multicolumn{2}{c|}{191.6} &
\multicolumn{2}{c|}{195.5} & \multicolumn{2}{c|}{199.5} &
\multicolumn{2}{c||}{201.6} &\multicolumn{2}{c|}{All} \\
Selection & 
Data & SM & Data & SM &  Data & SM & Data & SM & Data & SM & Data & SM \\
\hline
Four Jets &
126 & 133 & 19 & 22 & 57 & 57 & 61 & 60 & 23 & 29 & 
286 & 299\\
 Broad &&&&&&&&&&&& \\
\hline
Many Jets &
10 &  10 & 1 & 1.6 & 6 & 4.4 & 8 & 4.2 & 3 & 2.0 & 
28 & 22\\
&&&&&&&&&&&& \\
\hline
Many Jets + &
22 &  17 &3 & 2.9 & 8 & 7.0 & 10 & 7.8 & 9 & 3.8 &
52 & 39\\
Leptons   &&&&&&&&&&&& \\
\hline
Four Jets + &
 6 &   4.0 & 1 & 0.98 & 0 & 2.2 & 5 & 2.5 & 2 & 1.2 & 
14 & 11\\
2 Leptons   &&&&&&&&&&&& \\
\hline
Many Jets + &
 6 &   6.5 &2 & 1.4 & 2 & 2.8 & 2 & 4.0 & 3 & 1.9 &
 15 & 17\\
2 leptons   &&&&&&&&&&&& \\
\hline
Four Jets+$\emiss$ &
68 & 77 &15 & 14 & 35 & 34 & 40 & 37 & 20 & 18 &
178 & 175\\
&&&&&&&&&&&& \\

\hline
Many Jets+$\emiss$ &
87 &  80 &17 & 14 & 50 & 34 & 38 & 36 & 18 & 17 &
210 & 179\\
&&&&&&&&&&&& \\
\hline

\end{tabular}
\end{center}
\end{table}

\subsection{Charginos and neutralinos decaying via \boldmath ${\bar U}{\bar D}{\bar D}$ }
\label{udd.charginos}

The decay of the lightest neutralino and direct decays of the chargino
both lead to six hadronic jets in the final state.  The indirect
chargino decays give rise to a variety of final states depending on
the $\mathrm{W}^*$ decay, they range from ten hadronic jets to six
jets associated with leptons and missing energy.  The ``Many Jets'',
``Four Jets'' and ``Many Jets + Lepton'' selections are used to cover
these topologies.

Interpreting these results in the framework
of the MSSM, Fig.~\ref{mum2lims}(c) shows the $95\%$ C.L. exclusion in
the ($\mu$, $M_2$) plane. As for the $\slle$ and $\slqd$ couplings the
lower limit on the lightest chargino mass is essentially at the
kinematic limit.

The searches for the lightest and second lightest
neutralino do not extend the excluded region in the
($\mu$,$M_2$) plane beyond that achieved with the chargino 
search alone.

\subsection {Squarks decaying via \boldmath ${\bar U}{\bar D}{\bar D}$  }

The direct decay of pair produced squarks leads to four-quark final
states. The ``Four Jet'' selection is therefore used to extract the
mass limits.  As shown in Fig.~\ref{fig:fourjets}(b) the mass limits
are $82~\gevcc$ for up-type squarks and $68~\gevcc$ for down-type
squarks.

\begin{figure}
\begin{center}
\resizebox{\textwidth}{!}{
 \includegraphics{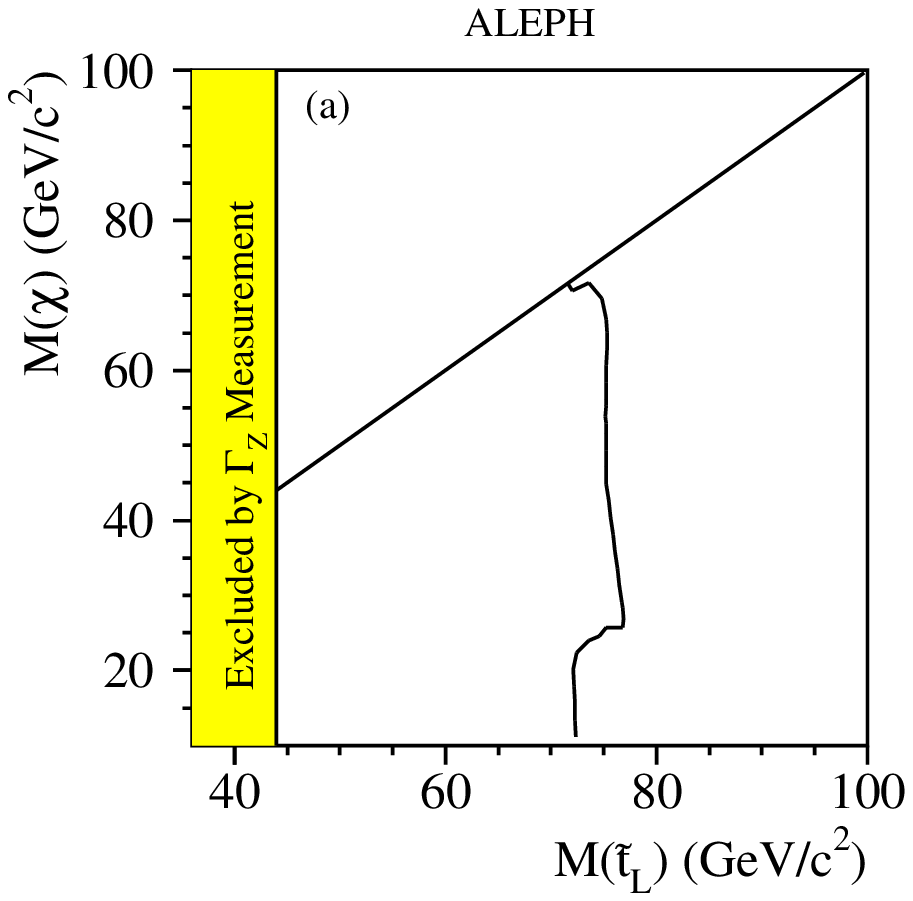}
 \includegraphics{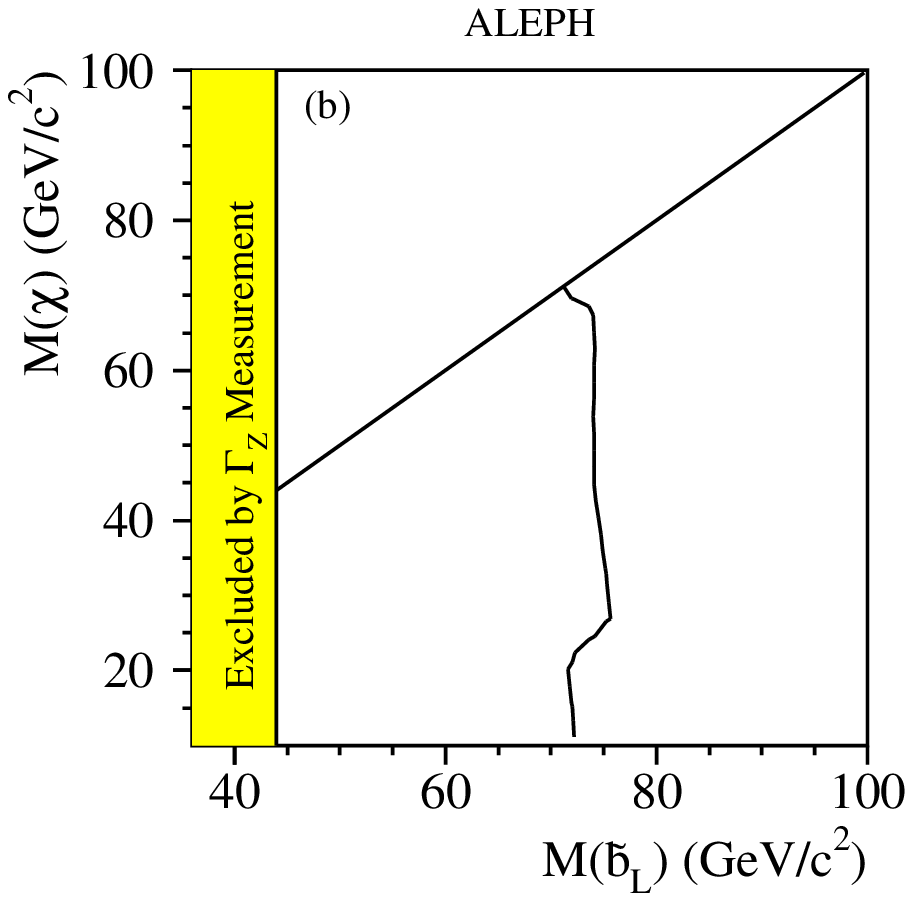}
}
\caption[.]{\label{udd_stop} \small The $95\%$ C.L. exclusion
in the ($M_{\chi}$, $M_{\sq}$) plane for
a) left-handed stop, b) left-handed sbottom
decaying indirectly via a dominant $\sudd$ operator.}
\end{center}
\end{figure}

For the indirect squark decays, which lead to eight-jet
topologies, the ``Four Jets-Broad'' selection is used. 
The resulting $95\%$
C.L. exclusion in the ($M_{\chi}$, $M_{\sq}$) plane for
left-handed stop and sbottom are shown in Fig.~\ref{udd_stop}.  The
corresponding mass limits are $M_{\stL}>71.5~\gevcc$ and
$M_{\sbL}>71.5~\gevcc$.

\subsection {Sleptons decaying via \boldmath ${\bar U}{\bar D}{\bar D}$  }

No direct slepton decays are possible via the $\sudd$ coupling.  For
the indirect decays of selectron and smuon pairs, which lead to
six-jets plus two-lepton final states, the ``Four Jets + 2 Leptons''
selection is used for large mass differences between the slepton and
neutralino, and the ``Many Jets + 2 Leptons'' for the low mass
difference region. In addition, for the very low mass difference
region the leptons are very soft and the ``Four Jets'' selection is
used.

\begin{figure}
\begin{center}
\resizebox{\textwidth}{!}{
  \includegraphics{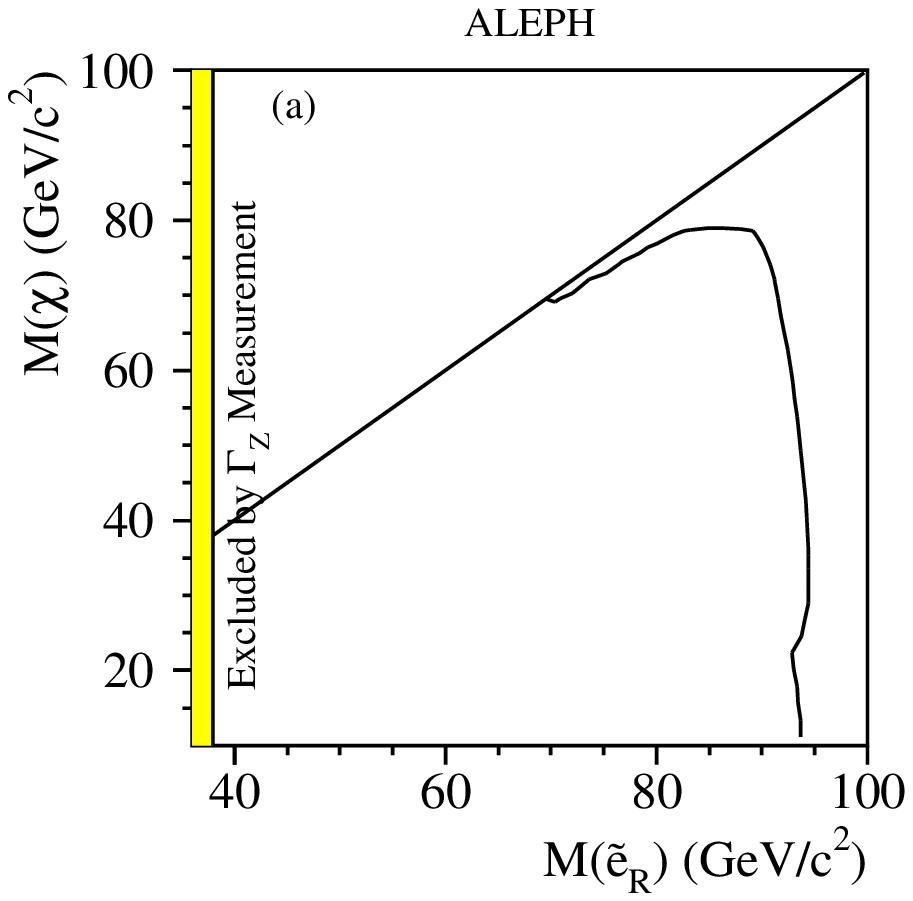}
  \includegraphics{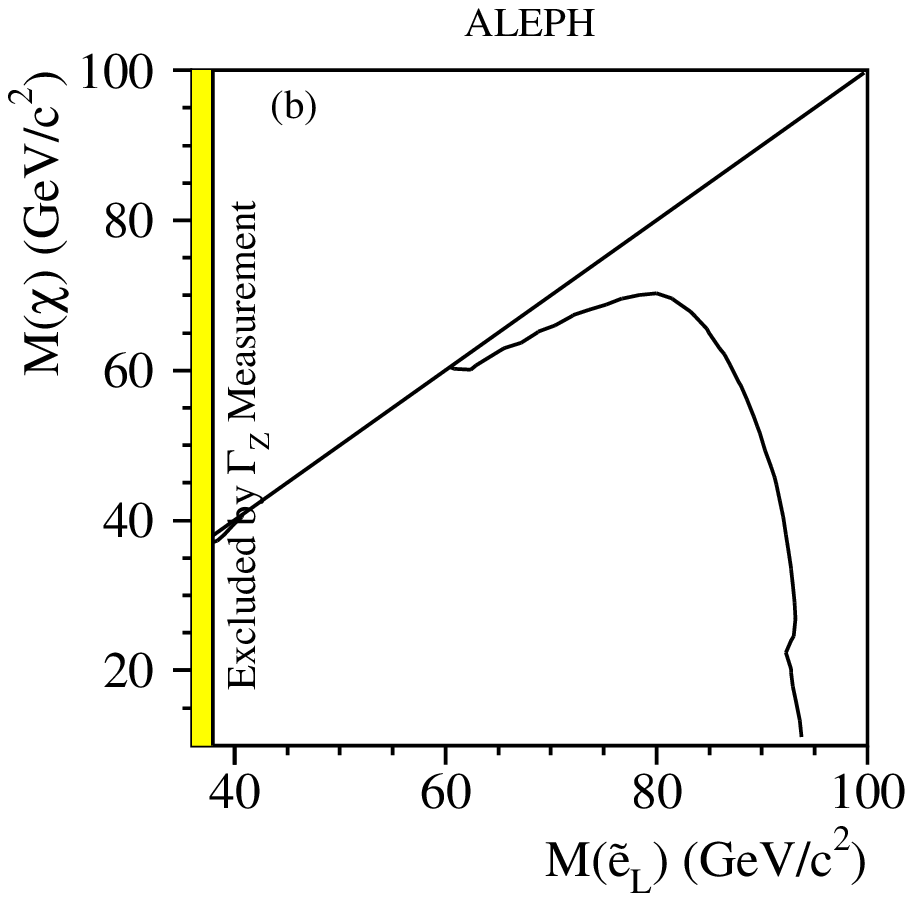}
}
\resizebox{\textwidth}{!}{
 \includegraphics{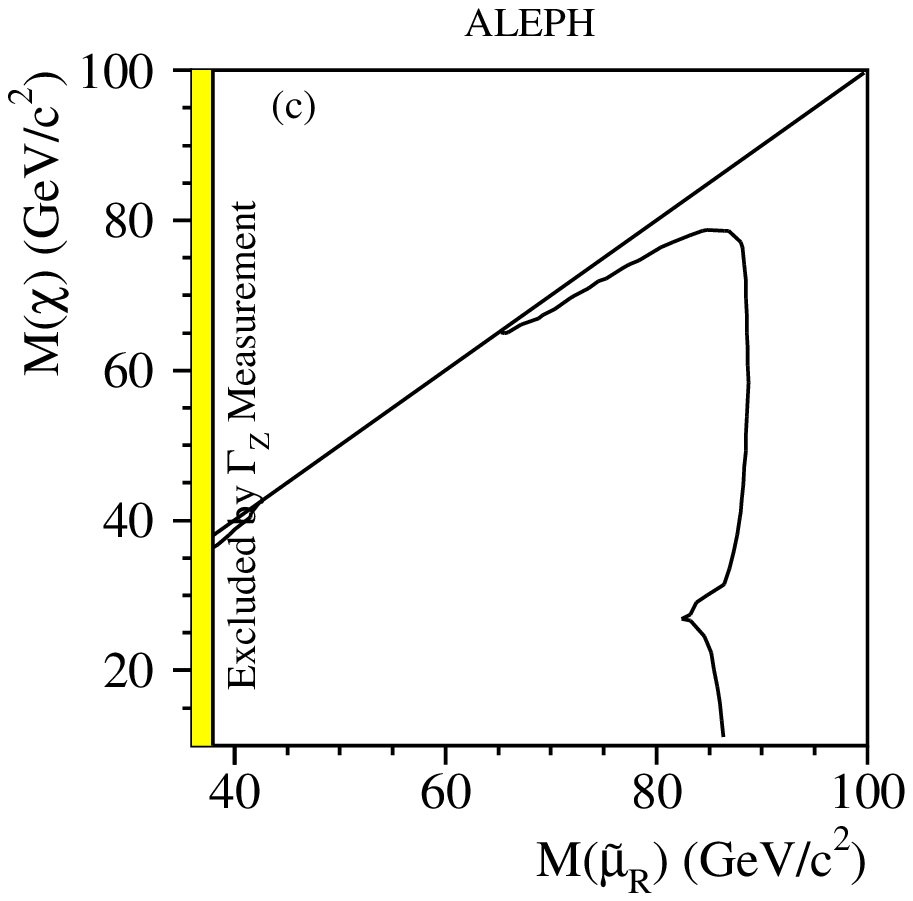}
 \includegraphics{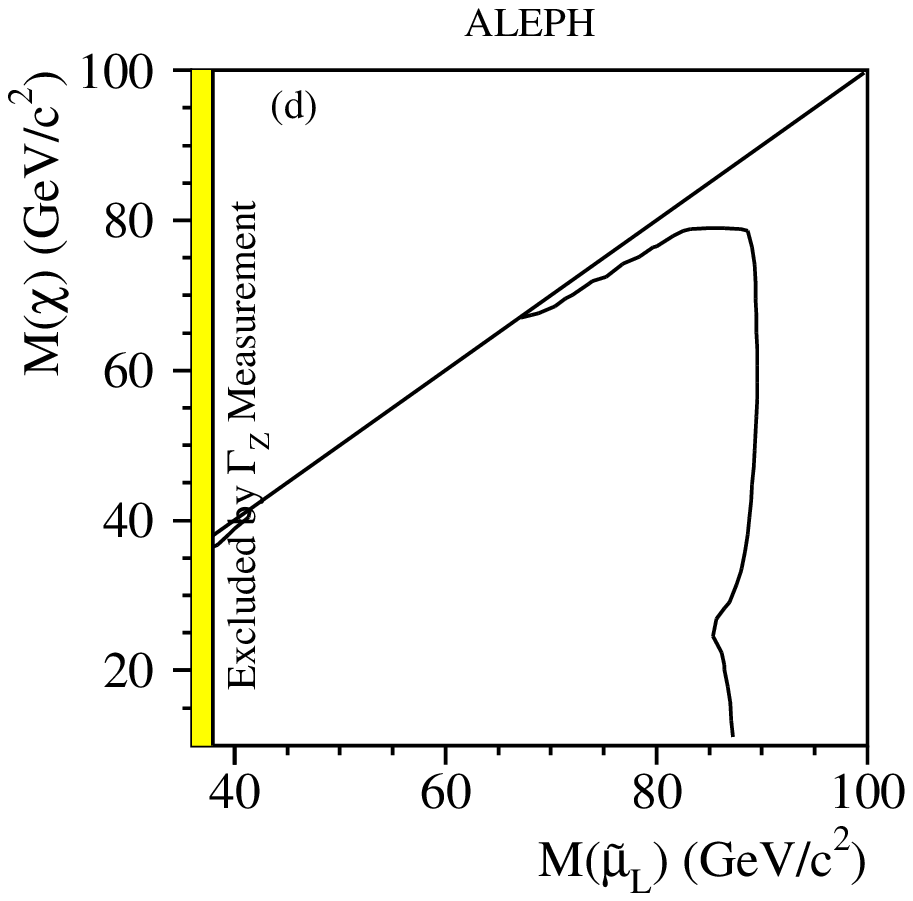}
}
\caption[.]{  \label{udd_slep}{\small The $95\%$ C.L. excluded
cross sections for left or right-handed selectrons and smuons decaying
indirectly via a dominant $\sudd$ operator. The selectron
cross section is evaluated in the region $\mu=-200~\gevcc$ and $\tanb=2$.}}
\end{center}
\end{figure}

Figure~\ref{udd_slep} shows the $95\%$ C.L. exclusion in the
($M_{\chi}$, $M_{\slep}$) plane for selectrons and smuons.  The
selectron cross section is evaluated at $\mu=-200~\gevcc$ and
$\tanb=2$.  The shape of the limits at $M_{\chi}\approx 20~\gevcc$ is
due to the switch between selections. For
$M_{\slep}-M_{\chi}>10~\gevcc$ this yields $M_{\seR}>88.5~\gevcc$ and
$M_{\smuR}>82.5~\gevcc$.

\subsection {Sneutrinos decaying via \boldmath ${\bar U}{\bar D}{\bar D}$  }
 \label{udd.sneutrinos}

No direct sneutrino decays are possible via the $\sudd$
coupling.  Sneutrinos decaying indirectly lead to six-jet final 
states containing two neutrinos. Large mass differences between the sneutrino
and neutralino lead to event topologies with significant missing
energy, and the ``Four Jets + $\emiss$'' selection is used.  For small mass
differences the ``Many Jets + $\emiss$'' selection is used.

\begin{figure}
\begin{center}
\resizebox{\textwidth}{!}{
  \includegraphics{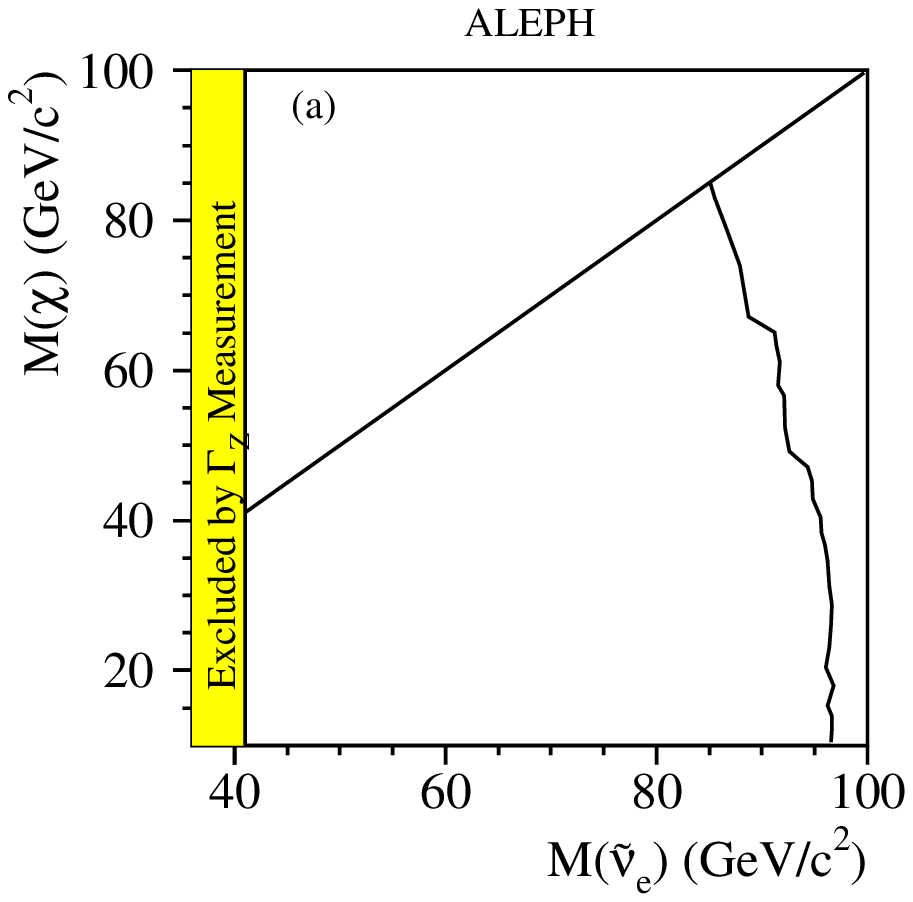}
  \includegraphics{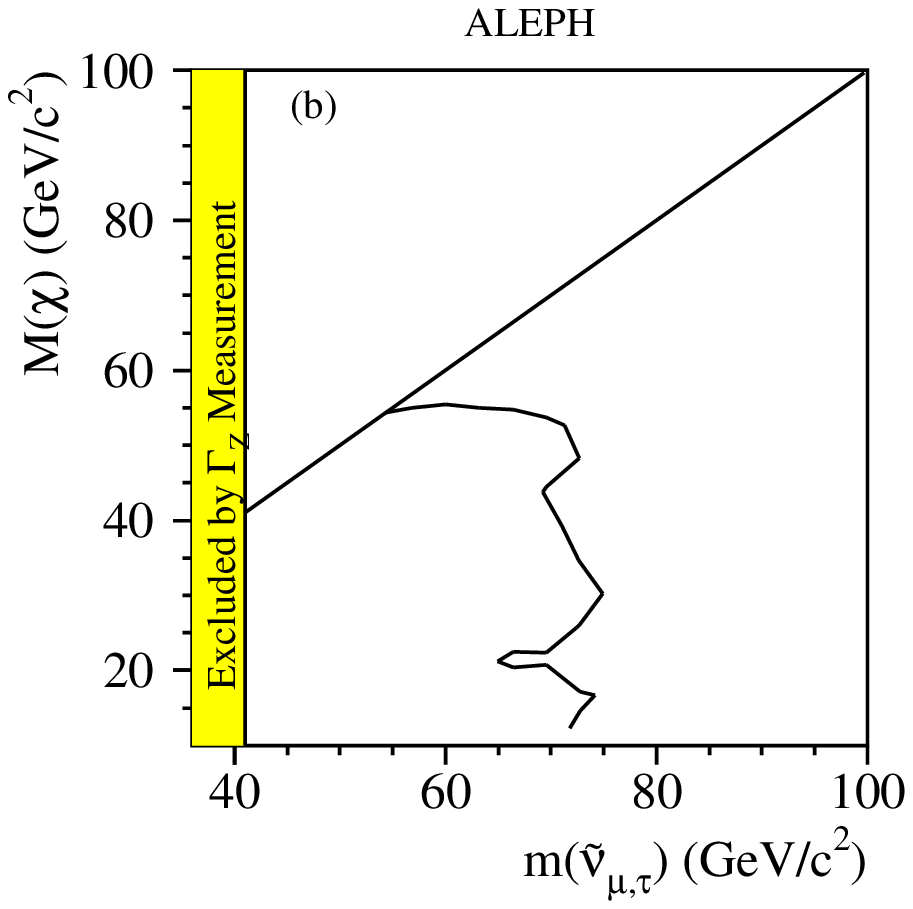}
}
\caption[.]{  \label{udd_sneu}{(a) \small The $95\%$ C.L. exclusion in the
($M_{\chi}$, $M_{\snu}$) plane for $\snu_e$ decaying indirectly via a
dominant $\sudd$ operator.  The $\snu_e$ cross section is evaluated at
$\mu=-200~\gevcc$ and $\tanb=2$. (b) The exclusion obtained in the
($M_{\chi}$, $M_{\snu_{\mu,\tau}}$) plane for $\snu_{\mu,\tau}$
decaying indirectly via a dominant $\sudd$ operator.}}
\end{center}
\end{figure}

Figure~\ref{udd_sneu} shows the $95\%$ C.L. exclusion in the
($M_{\chi}$, $M_{\snu}$) plane for the electron sneutrino
(Fig.~\ref{udd_sneu}(a)) and muon or tau sneutrino
(Fig.~\ref{udd_sneu}(b)). The electron sneutrino cross sections is
evaluated at $\mu=-200~\gevcc$ and $\tanb=2$. The limits
$M_{\snue}>84~\gevcc$ and $M_{\snu_{\mu,\tau}}>64~\gevcc$ are
obtained.

\section{Summary}\label{conclusions}
A number of searches were developed to select R-parity violating decay
topologies for single and pair production of SUSY particles.  It has been
assumed that the LSP has a negligible lifetime, and that only one
coupling $\lambda_{ijk},\lambda'_{ijk}$ or $\lambda''_{ijk}$ is
nonzero.  These searches found no evidence for R-parity violating
supersymmetry in the data collected at $\sqrt{s}=188.6$--$201.6~\gev$,
and various limits were set within the framework of the MSSM.

From searches for singly produced sneutrinos,
upper limits on the values of the $\lambda_{121}$ and $\lambda_{131}$
couplings were set as a function of $M_{\snu_{\mu}}$ and
$M_{\snu_{\tau}}$, respectively.
 
The limits for direct decays of sleptons for an $\slle$ coupling are
$M_{\seR}>92~\gevcc$ and $M_{\snue}>98~\gevcc$ at $\mu=-200~\gevcc$
and $\tanb=2$, $M_{\smuR,\stauR}>81~\gevcc$ and
$M_{\snu_{\mu,\tau}}>86~\gevcc$. The limits for the direct decays of
sleptons and squarks in the case of an $\slqd$ coupling are
$M_{\smuL}>79~\gevcc$, $M_{\snu_\mu}>77~\gevcc$ and
$M_{\stL}>93~\gevcc$ for Br$(\stL\ra \mathrm{q}\tau)=1$. The limit for
squarks assuming a $\sudd$ coupling are $M_{\suR}>82~\gevcc$ and
$M_{\sdR}>68~\gevcc$.

For the indirect decays of sfermions, the limits listed in
Table~\ref{table:indirect_limits} have been
obtained, assuming $M_{\slep}-M_{\chi}>10~\gevcc$ for $\slqd$ and
$\sudd$, $M_{\chi}>23~\gevcc$ for $\slle$ and derived at
$\mu=-200~\gevcc$ and $\tanb=2$ for $\se$ and $\snue$.

\begin{table}
\caption[.]{ \small 
The $95\%$ confidence level lower mass limits for sparticles decaying
indirectly for each of the three R-parity violating couplings.
}
\label{table:indirect_limits}
\begin{center}
\renewcommand{\arraystretch}{1.1}
\begin{tabular}{|c|c|c|c|}
\hline
 & \multicolumn{3}{c|}{Lower mass limit ($\gevcc$)} \\
\cline{2-4} 
Sparticle &~~~$\slle$~~~ & ~~~$\slqd$~~~ & ~~~$\sudd$~~~ \\
\hline
$\stone$ & 90  & 84  & 71.5  \\
$\sbone$ & 89  & 74  & 71.5  \\
$\seR$   & 93  & 89  & 88.5  \\
$\smuR$  & 92  & 86  & 82.5  \\
$\stauR$ & 91  & 73  & $\times$ \\
$\snue$  & 94  & 89  & 84  \\
$\snu_{\mu,\tau}$ &  83  & 75  & 64 \\
\hline
\end{tabular}
\renewcommand{\arraystretch}{1.}
\end{center}
\end{table}

Assuming large $m_0$ the chargino mass limit is given by the kinematic
limit $M_{\chi^+}>100~\gevcc$, irrespective of the R-parity violating
operator.

\section{Acknowledgements}
It is a pleasure to congratulate our colleagues from the accelerator divisions
for the successful operation of LEP at high energy. 
We would like to express our gratitude to the engineers and 
support people at our home institutes without whose dedicated help
this work would not have been possible. 
Those of us from non-member states wish to thank CERN for its hospitality
and support.

\clearpage

\end{document}

%% file: authb.tex
\pagestyle{empty}
\newpage
\small
%
%
\newlength{\saveparskip}
\newlength{\savetextheight}
\newlength{\savetopmargin}
\newlength{\savetextwidth}
\newlength{\saveoddsidemargin}
\newlength{\savetopsep}
\setlength{\saveparskip}{\parskip}
\setlength{\savetextheight}{\textheight}
\setlength{\savetopmargin}{\topmargin}
\setlength{\savetextwidth}{\textwidth}
\setlength{\saveoddsidemargin}{\oddsidemargin}
\setlength{\savetopsep}{\topsep}
%
%
\setlength{\parskip}{0.0cm}
\setlength{\textheight}{25.0cm}
\setlength{\topmargin}{-1.5cm}
\setlength{\textwidth}{16 cm}
\setlength{\oddsidemargin}{-0.0cm}
\setlength{\topsep}{1mm}
\pretolerance=10000
\centerline{\large\bf The ALEPH Collaboration}
\footnotesize
\vspace{0.5cm}
{\raggedbottom
\begin{sloppypar}
\samepage\noindent
R.~Barate,
D.~Decamp,
P.~Ghez,
C.~Goy,
S.~Jezequel,
J.-P.~Lees,
F.~Martin,
E.~Merle,
\mbox{M.-N.~Minard},
B.~Pietrzyk
\nopagebreak
\begin{center}
\parbox{15.5cm}{\sl\samepage
Laboratoire de Physique des Particules (LAPP), IN$^{2}$P$^{3}$-CNRS,
F-74019 Annecy-le-Vieux Cedex, France}
\end{center}\end{sloppypar}
\vspace{2mm}
\begin{sloppypar}
\noindent
S.~Bravo,
M.P.~Casado,
M.~Chmeissani,
J.M.~Crespo,
E.~Fernandez,
M.~Fernandez-Bosman,
Ll.~Garrido,$^{15}$
E.~Graug\'{e}s,
J.~Lopez,
M.~Martinez,
G.~Merino,
R.~Miquel,
Ll.M.~Mir,
A.~Pacheco,
D.~Paneque,
H.~Ruiz
\nopagebreak
\begin{center}
\parbox{15.5cm}{\sl\samepage
Institut de F\'{i}sica d'Altes Energies, Universitat Aut\`{o}noma
de Barcelona, E-08193 Bellaterra (Barcelona), Spain$^{7}$}
\end{center}\end{sloppypar}
\vspace{2mm}
\begin{sloppypar}
\noindent
A.~Colaleo,
D.~Creanza,
N.~De~Filippis,
M.~de~Palma,
G.~Iaselli,
G.~Maggi,
M.~Maggi,$^{1}$
S.~Nuzzo,
A.~Ranieri,
G.~Raso,$^{24}$
F.~Ruggieri,
G.~Selvaggi,
L.~Silvestris,
P.~Tempesta,
A.~Tricomi,$^{3}$
G.~Zito
\nopagebreak
\begin{center}
\parbox{15.5cm}{\sl\samepage
Dipartimento di Fisica, INFN Sezione di Bari, I-70126 Bari, Italy}
\end{center}\end{sloppypar}
\vspace{2mm}
\begin{sloppypar}
\noindent
X.~Huang,
J.~Lin,
Q. Ouyang,
T.~Wang,
Y.~Xie,
R.~Xu,
S.~Xue,
J.~Zhang,
L.~Zhang,
W.~Zhao
\nopagebreak
\begin{center}
\parbox{15.5cm}{\sl\samepage
Institute of High Energy Physics, Academia Sinica, Beijing, The People's
Republic of China$^{8}$}
\end{center}\end{sloppypar}
\vspace{2mm}
\begin{sloppypar}
\noindent
D.~Abbaneo,
P.~Azzurri,
G.~Boix,$^{6}$
O.~Buchm\"uller,
M.~Cattaneo,
F.~Cerutti,
B.~Clerbaux,
G.~Dissertori,
H.~Drevermann,
R.W.~Forty,
M.~Frank,
F.~Gianotti,
T.C.~Greening,
J.B.~Hansen,
J.~Harvey,
D.E.~Hutchcroft,
P.~Janot,
B.~Jost,
M.~Kado,
V.~Lemaitre,
P.~Maley,
P.~Mato,
A.~Minten,
A.~Moutoussi,
F.~Ranjard,
L.~Rolandi,
D.~Schlatter,
M.~Schmitt,$^{20}$
O.~Schneider,$^{2}$
P.~Spagnolo,
W.~Tejessy,
F.~Teubert,
E.~Tournefier,$^{26}$
A.~Valassi,
J.J.~Ward,
A.E.~Wright
\nopagebreak
\begin{center}
\parbox{15.5cm}{\sl\samepage
European Laboratory for Particle Physics (CERN), CH-1211 Geneva 23,
Switzerland}
\end{center}\end{sloppypar}
\vspace{2mm}
\begin{sloppypar}
\noindent
Z.~Ajaltouni,
F.~Badaud,
S.~Dessagne,
A.~Falvard,
D.~Fayolle,
P.~Gay,
P.~Henrard,
J.~Jousset,
B.~Michel,
S.~Monteil,
\mbox{J-C.~Montret},
D.~Pallin,
J.M.~Pascolo,
P.~Perret,
F.~Podlyski
\nopagebreak
\begin{center}
\parbox{15.5cm}{\sl\samepage
Laboratoire de Physique Corpusculaire, Universit\'e Blaise Pascal,
IN$^{2}$P$^{3}$-CNRS, Clermont-Ferrand, F-63177 Aubi\`{e}re, France}
\end{center}\end{sloppypar}
\vspace{2mm}
\begin{sloppypar}
\noindent
J.D.~Hansen,
J.R.~Hansen,
P.H.~Hansen,
B.S.~Nilsson,
A.~W\"a\"an\"anen
\nopagebreak
\begin{center}
\parbox{15.5cm}{\sl\samepage
Niels Bohr Institute, 2100 Copenhagen, DK-Denmark$^{9}$}
\end{center}\end{sloppypar}
\vspace{2mm}
\begin{sloppypar}
\noindent
G.~Daskalakis,
A.~Kyriakis,
C.~Markou,
E.~Simopoulou,
A.~Vayaki
\nopagebreak
\begin{center}
\parbox{15.5cm}{\sl\samepage
Nuclear Research Center Demokritos (NRCD), GR-15310 Attiki, Greece}
\end{center}\end{sloppypar}
\vspace{2mm}
\begin{sloppypar}
\noindent
A.~Blondel,$^{12}$
\mbox{J.-C.~Brient},
F.~Machefert,
A.~Roug\'{e},
M.~Swynghedauw,
R.~Tanaka
\linebreak
H.~Videau
\nopagebreak
\begin{center}
\parbox{15.5cm}{\sl\samepage
Laboratoire de Physique Nucl\'eaire et des Hautes Energies, Ecole
Polytechnique, IN$^{2}$P$^{3}$-CNRS, \mbox{F-91128} Palaiseau Cedex, France}
\end{center}\end{sloppypar}
\vspace{2mm}
\begin{sloppypar}
\noindent
E.~Focardi,
G.~Parrini,
K.~Zachariadou
\nopagebreak
\begin{center}
\parbox{15.5cm}{\sl\samepage
Dipartimento di Fisica, Universit\`a di Firenze, INFN Sezione di Firenze,
I-50125 Firenze, Italy}
\end{center}\end{sloppypar}
\vspace{2mm}
\begin{sloppypar}
\noindent
A.~Antonelli,
M.~Antonelli,
G.~Bencivenni,
G.~Bologna,$^{4}$
F.~Bossi,
P.~Campana,
G.~Capon,
V.~Chiarella,
P.~Laurelli,
G.~Mannocchi,$^{5}$
F.~Murtas,
G.P.~Murtas,
L.~Passalacqua,
M.~Pepe-Altarelli$^{25}$
\nopagebreak
\begin{center}
\parbox{15.5cm}{\sl\samepage
Laboratori Nazionali dell'INFN (LNF-INFN), I-00044 Frascati, Italy}
\end{center}\end{sloppypar}
\vspace{2mm}
\begin{sloppypar}
\noindent
M.~Chalmers,
A.W.~Halley,
J.~Kennedy,
J.G.~Lynch,
P.~Negus,
V.~O'Shea,
B.~Raeven,
D.~Smith,
P.~Teixeira-Dias,
A.S.~Thompson
\nopagebreak
\begin{center}
\parbox{15.5cm}{\sl\samepage
Department of Physics and Astronomy, University of Glasgow, Glasgow G12
8QQ,United Kingdom$^{10}$}
\end{center}\end{sloppypar}
\begin{sloppypar}
\noindent
R.~Cavanaugh,
S.~Dhamotharan,
C.~Geweniger,
P.~Hanke,
V.~Hepp,
E.E.~Kluge,
G.~Leibenguth,
A.~Putzer,
K.~Tittel,
S.~Werner,$^{19}$
M.~Wunsch$^{19}$
\nopagebreak
\begin{center}
\parbox{15.5cm}{\sl\samepage
Kirchhoff-Institut f\"ur Physik, Universit\"at Heidelberg, D-69120
Heidelberg, Germany$^{16}$}
\end{center}\end{sloppypar}
\vspace{2mm}
\begin{sloppypar}
\noindent
R.~Beuselinck,
D.M.~Binnie,
W.~Cameron,
G.~Davies,
P.J.~Dornan,
M.~Girone,$^{1}$
N.~Marinelli,
J.~Nowell,
H.~Przysiezniak,
J.K.~Sedgbeer,
J.C.~Thompson,$^{14}$
E.~Thomson,$^{23}$
R.~White
\nopagebreak
\begin{center}
\parbox{15.5cm}{\sl\samepage
Department of Physics, Imperial College, London SW7 2BZ,
United Kingdom$^{10}$}
\end{center}\end{sloppypar}
\vspace{2mm}
\begin{sloppypar}
\noindent
V.M.~Ghete,
P.~Girtler,
E.~Kneringer,
D.~Kuhn,
G.~Rudolph
\nopagebreak
\begin{center}
\parbox{15.5cm}{\sl\samepage
Institut f\"ur Experimentalphysik, Universit\"at Innsbruck, A-6020
Innsbruck, Austria$^{18}$}
\end{center}\end{sloppypar}
\vspace{2mm}
\begin{sloppypar}
\noindent
E.~Bouhova-Thacker,
C.K.~Bowdery,
D.P.~Clarke,
G.~Ellis,
A.J.~Finch,
F.~Foster,
G.~Hughes,
R.W.L.~Jones,$^{1}$
M.R.~Pearson,
N.A.~Robertson,
M.~Smizanska
\nopagebreak
\begin{center}
\parbox{15.5cm}{\sl\samepage
Department of Physics, University of Lancaster, Lancaster LA1 4YB,
United Kingdom$^{10}$}
\end{center}\end{sloppypar}
\vspace{2mm}
\begin{sloppypar}
\noindent
I.~Giehl,
F.~H\"olldorfer,
K.~Jakobs,
K.~Kleinknecht,
M.~Kr\"ocker,
A.-S.~M\"uller,
H.-A.~N\"urnberger,
G.~Quast,$^{1}$
B.~Renk,
E.~Rohne,
H.-G.~Sander,
S.~Schmeling,
H.~Wachsmuth,
C.~Zeitnitz,
T.~Ziegler
\nopagebreak
\begin{center}
\parbox{15.5cm}{\sl\samepage
Institut f\"ur Physik, Universit\"at Mainz, D-55099 Mainz, Germany$^{16}$}
\end{center}\end{sloppypar}
\vspace{2mm}
\begin{sloppypar}
\noindent
A.~Bonissent,
J.~Carr,
P.~Coyle,
C.~Curtil,
A.~Ealet,
D.~Fouchez,
O.~Leroy,
T.~Kachelhoffer,
P.~Payre,
D.~Rousseau,
A.~Tilquin
\nopagebreak
\begin{center}
\parbox{15.5cm}{\sl\samepage
Centre de Physique des Particules de Marseille, Univ M\'editerran\'ee,
IN$^{2}$P$^{3}$-CNRS, F-13288 Marseille, France}
\end{center}\end{sloppypar}
\vspace{2mm}
\begin{sloppypar}
\noindent
M.~Aleppo,
S.~Gilardoni,
F.~Ragusa
\nopagebreak
\begin{center}
\parbox{15.5cm}{\sl\samepage
Dipartimento di Fisica, Universit\`a di Milano e INFN Sezione di
Milano, I-20133 Milano, Italy.}
\end{center}\end{sloppypar}
\vspace{2mm}
\begin{sloppypar}
\noindent
A.~David,
H.~Dietl,
G.~Ganis,$^{27}$
A.~Heister,
K.~H\"uttmann,
G.~L\"utjens,
C.~Mannert,
W.~M\"anner,
\mbox{H.-G.~Moser},
S.~Schael,
R.~Settles,$^{1}$
H.~Stenzel,
W.~Wiedenmann,
G.~Wolf
\nopagebreak
\begin{center}
\parbox{15.5cm}{\sl\samepage
Max-Planck-Institut f\"ur Physik, Werner-Heisenberg-Institut,
D-80805 M\"unchen, Germany\footnotemark[16]}
\end{center}\end{sloppypar}
\vspace{2mm}
\begin{sloppypar}
\noindent
J.~Boucrot,$^{1}$
O.~Callot,
M.~Davier,
L.~Duflot,
\mbox{J.-F.~Grivaz},
Ph.~Heusse,
A.~Jacholkowska,$^{1}$
L.~Serin,
\mbox{J.-J.~Veillet},
I.~Videau,
J.-B.~de~Vivie~de~R\'egie,$^{28}$
C.~Yuan,
D.~Zerwas
\nopagebreak
\begin{center}
\parbox{15.5cm}{\sl\samepage
Laboratoire de l'Acc\'el\'erateur Lin\'eaire, Universit\'e de Paris-Sud,
IN$^{2}$P$^{3}$-CNRS, F-91898 Orsay Cedex, France}
\end{center}\end{sloppypar}
\vspace{2mm}
\begin{sloppypar}
\noindent
G.~Bagliesi,
T.~Boccali,
G.~Calderini,
V.~Ciulli,
L.~Fo\`a,
A.~Giammanco,
A.~Giassi,
F.~Ligabue,
A.~Messineo,
F.~Palla,$^{1}$
G.~Sanguinetti,
A.~Sciab\`a,
G.~Sguazzoni,
R.~Tenchini,$^{1}$
A.~Venturi,
P.G.~Verdini
\samepage
\begin{center}
\parbox{15.5cm}{\sl\samepage
Dipartimento di Fisica dell'Universit\`a, INFN Sezione di Pisa,
e Scuola Normale Superiore, I-56010 Pisa, Italy}
\end{center}\end{sloppypar}
\vspace{2mm}
\begin{sloppypar}
\noindent
G.A.~Blair,
J.~Coles,
G.~Cowan,
M.G.~Green,
L.T.~Jones,
T.~Medcalf,
J.A.~Strong,
\mbox{J.H.~von~Wimmersperg-Toeller} 
\nopagebreak
\begin{center}
\parbox{15.5cm}{\sl\samepage
Department of Physics, Royal Holloway \& Bedford New College,
University of London, Surrey TW20 OEX, United Kingdom$^{10}$}
\end{center}\end{sloppypar}
\vspace{2mm}
\begin{sloppypar}
\noindent
R.W.~Clifft,
T.R.~Edgecock,
P.R.~Norton,
I.R.~Tomalin
\nopagebreak
\begin{center}
\parbox{15.5cm}{\sl\samepage
Particle Physics Dept., Rutherford Appleton Laboratory,
Chilton, Didcot, Oxon OX11 OQX, United Kingdom$^{10}$}
\end{center}\end{sloppypar}
\vspace{2mm}
\begin{sloppypar}
\noindent
\mbox{B.~Bloch-Devaux},$^{1}$
D.~Boumediene,
P.~Colas,
B.~Fabbro,
E.~Lan\c{c}on,
\mbox{M.-C.~Lemaire},
E.~Locci,
P.~Perez,
J.~Rander,
\mbox{J.-F.~Renardy},
A.~Rosowsky,
P.~Seager,$^{13}$
A.~Trabelsi,$^{21}$
B.~Tuchming,
B.~Vallage
\nopagebreak
\begin{center}
\parbox{15.5cm}{\sl\samepage
CEA, DAPNIA/Service de Physique des Particules,
CE-Saclay, F-91191 Gif-sur-Yvette Cedex, France$^{17}$}
\end{center}\end{sloppypar}
\vspace{2mm}
\begin{sloppypar}
\noindent
N.~Konstantinidis,
C.~Loomis,
A.M.~Litke,
G.~Taylor
\nopagebreak
\begin{center}
\parbox{15.5cm}{\sl\samepage
Institute for Particle Physics, University of California at
Santa Cruz, Santa Cruz, CA 95064, USA$^{22}$}
\end{center}\end{sloppypar}
\vspace{2mm}
\begin{sloppypar}
\noindent
C.N.~Booth,
S.~Cartwright,
F.~Combley,
P.N.~Hodgson,
M.~Lehto,
L.F.~Thompson
\nopagebreak
\begin{center}
\parbox{15.5cm}{\sl\samepage
Department of Physics, University of Sheffield, Sheffield S3 7RH,
United Kingdom$^{10}$}
\end{center}\end{sloppypar}
\vspace{2mm}
\begin{sloppypar}
\noindent
K.~Affholderbach,
A.~B\"ohrer,
S.~Brandt,
C.~Grupen,
J.~Hess,
A.~Misiejuk,
G.~Prange,
U.~Sieler
\nopagebreak
\begin{center}
\parbox{15.5cm}{\sl\samepage
Fachbereich Physik, Universit\"at Siegen, D-57068 Siegen, Germany$^{16}$}
\end{center}\end{sloppypar}
\vspace{2mm}
\begin{sloppypar}
\noindent
C.~Borean,
G.~Giannini,
B.~Gobbo
\nopagebreak
\begin{center}
\parbox{15.5cm}{\sl\samepage
Dipartimento di Fisica, Universit\`a di Trieste e INFN Sezione di Trieste,
I-34127 Trieste, Italy}
\end{center}\end{sloppypar}
\vspace{2mm}
\begin{sloppypar}
\noindent
H.~He,
J.~Putz,
J.~Rothberg,
S.~Wasserbaech
\nopagebreak
\begin{center}
\parbox{15.5cm}{\sl\samepage
Experimental Elementary Particle Physics, University of Washington, Seattle,
WA 98195 U.S.A.}
\end{center}\end{sloppypar}
\vspace{2mm}
\begin{sloppypar}
\noindent
S.R.~Armstrong,
K.~Cranmer,
P.~Elmer,
D.P.S.~Ferguson,
Y.~Gao,
S.~Gonz\'{a}lez,
O.J.~Hayes,
H.~Hu,
S.~Jin,
J.~Kile,
P.A.~McNamara III,
J.~Nielsen,
W.~Orejudos,
Y.B.~Pan,
Y.~Saadi,
I.J.~Scott,
J.~Walsh,
J.~Wu,
Sau~Lan~Wu,
X.~Wu,
G.~Zobernig
\nopagebreak
\begin{center}
\parbox{15.5cm}{\sl\samepage
Department of Physics, University of Wisconsin, Madison, WI 53706,
USA$^{11}$}
\end{center}\end{sloppypar}
}
\footnotetext[1]{Also at CERN, 1211 Geneva 23, Switzerland.}
\footnotetext[2]{Now at Universit\'e de Lausanne, 1015 Lausanne, Switzerland.}
\footnotetext[3]{Also at Dipartimento di Fisica di Catania and INFN Sezione di
 Catania, 95129 Catania, Italy.}
\footnotetext[4]{Deceased.}
\footnotetext[5]{Also Istituto di Cosmo-Geofisica del C.N.R., Torino,
Italy.}
\footnotetext[6]{Supported by the Commission of the European Communities,
contract ERBFMBICT982894.}
\footnotetext[7]{Supported by CICYT, Spain.}
\footnotetext[8]{Supported by the National Science Foundation of China.}
\footnotetext[9]{Supported by the Danish Natural Science Research Council.}
\footnotetext[10]{Supported by the UK Particle Physics and Astronomy Research
Council.}
\footnotetext[11]{Supported by the US Department of Energy, grant
DE-FG0295-ER40896.}
\footnotetext[12]{Now at Departement de Physique Corpusculaire, Universit\'e de
Gen\`eve, 1211 Gen\`eve 4, Switzerland.}
\footnotetext[13]{Supported by the Commission of the European Communities,
contract ERBFMBICT982874.}
\footnotetext[14]{Also at Rutherford Appleton Laboratory, Chilton, Didcot, UK.}
\footnotetext[15]{Permanent address: Universitat de Barcelona, 08208 Barcelona,
Spain.}
\footnotetext[16]{Supported by the Bundesministerium f\"ur Bildung,
Wissenschaft, Forschung und Technologie, Germany.}
\footnotetext[17]{Supported by the Direction des Sciences de la
Mati\`ere, C.E.A.}
\footnotetext[18]{Supported by the Austrian Ministry for Science and Transport.}
\footnotetext[19]{Now at SAP AG, 69185 Walldorf, Germany}
\footnotetext[20]{Now at Harvard University, Cambridge, MA 02138, U.S.A.}
\footnotetext[21]{Now at D\'epartement de Physique, Facult\'e des Sciences de Tunis, 1060 Le Belv\'ed\`ere, Tunisia.}
\footnotetext[22]{Supported by the US Department of Energy,
grant DE-FG03-92ER40689.}
\footnotetext[23]{Now at Department of Physics, Ohio State University, Columbus, OH 43210-1106, U.S.A.}
\footnotetext[24]{Also at Dipartimento di Fisica e Tecnologie Relative, Universit\`a di Palermo, Palermo, Italy.}
\footnotetext[25]{Now at CERN, 1211 Geneva 23, Switzerland.}
\footnotetext[26]{Now at ISN, Institut des Sciences Nucl\'eaires, 53 Av. des Martyrs, 38026 Grenoble, France.}
\footnotetext[27]{Now at Universit\`a degli Studi di Roma Tor Vergata, Dipartimento di Fisica, 00133 Roma, Italy.}
\footnotetext[28]{Now at Centre de Physique des Particules de Marseille,Univ M\'editerran\'ee, F-13288 Marseille, France.}
%
\setlength{\parskip}{\saveparskip}
\setlength{\textheight}{\savetextheight}
\setlength{\topmargin}{\savetopmargin}
\setlength{\textwidth}{\savetextwidth}
\setlength{\oddsidemargin}{\saveoddsidemargin}
\setlength{\topsep}{\savetopsep}
\normalsize
\newpage
\pagestyle{plain}
\setcounter{page}{1}